\documentclass[12pt,preprint]{aastex}

\shorttitle{IRAC OBSERVATIONS OF WHITE DWARFS. I.}
\shortauthors{J. Farihi}

\begin{document}

\title{{\em SPITZER} IRAC OBSERVATIONS OF WHITE DWARFS.\\
	I. WARM DUST AT METAL-RICH DEGENERATES}

\author{J. Farihi\altaffilmark{1,2},
	 B. Zuckerman\altaffilmark{1}, \& 
	 E. E. Becklin\altaffilmark{1}}

\altaffiltext{1}{Department of Physics \& Astronomy,
			University of California,
			430 Portola Plaza,
			Los Angeles, CA 90095; jfarihi,ben,becklin@astro.ucla.edu}
			
\altaffiltext{2}{Gemini Observatory,
			Northern Operations,
			670 North A'ohoku Place,
			Hilo, HI 96720}

\begin{abstract}

This paper presents the results of a {\em Spitzer} IRAC $3-8$ $\mu$m 
photometric search for warm dust orbiting 17 nearby, metal-rich white
dwarfs, 15 of which apparently have hydrogen dominated atmospheres 
(type DAZ).  G166-58, G29-38, and GD 362 manifest excess emission in 
their IRAC fluxes and the latter two are known to harbor dust grains warm 
enough to radiate detectable emission at near-infrared wavelengths as 
short as 2 $\mu$m.  Their IRAC fluxes display differences compatible with 
a relatively larger amount of cooler dust at GD 362.  G166-58 is presently 
unique in that it appears to exhibit excess flux only at wavelengths longer 
than about 5 $\mu$m.  Evidence is presented that this mid-infrared emission
is most likely associated with the white dwarf, indicating that G166-58 bears
circumstellar dust no warmer than $T\sim400$ K.  The remaining 14 targets 
reveal no reliable mid-infrared excess, indicating the majority of DAZ stars 
do not have warm debris disks sufficiently opaque to be detected  by IRAC.

\end{abstract}

\keywords{circumstellar matter---
	infrared: stars--
	minor planets, asteroids---
	planetary systems --
	stars: abundances---
	stars: evolution---
	stars: individual (G166-58, G29-38, GD 362, PG 0235$+$064)---
	white dwarfs}

\section{INTRODUCTION}

The {\em Spitzer Space Telescope} opened a new phase space to white dwarf
researchers interested in the infrared properties of degenerate stars and 
their environments.  The majority of white dwarfs are inaccessible from the
ground beyond 2.4 $\mu$m due to their intrinsic faintness combined with the 
ever-increasing sky brightness towards longer wavelengths \citep*{gla99}.
This limits any white dwarf science which aims to study matter radiating 
at $T<1500$ K.  Prior to the launch of {\em Spitzer}, only one previously 
published, directed mid-infrared study of white dwarfs exists ; an {\em 
Infrared Space Observatory} search for dust emission around 11 nearby
white dwarfs, 6 of which have metal-rich photospheres \citep*{cha99}.

Owing to the superb sensitivity of {\em Spitzer} \citep*{wer04}, a
Cycle 1 IRAC program was undertaken to search for warm dust emission 
associated with cool, hydrogen atmosphere white dwarfs with photospheric
metals, the DAZ stars.  This paper presents a synopsis of the IRAC results,
including the detection of $5-8$ $\mu$m flux excess at G166-58, $3-8$ $\mu$m 
data on G29-38 and GD 362, and also includes Gemini $3-4$ $\mu$m spectroscopy
of G29-38.

\section{SCIENTIFIC MOTIVATION}

\subsection{White Dwarfs Are Metal-Poor}

The origin and abundances of photospheric metals in isolated white
dwarfs has been an astrophysical curiosity dating back to the era 
when the first few white dwarfs were finally understood to be
subluminous via the combination of spectra and parallax; Sirius B,
40 Eri B, and van Maanen 2 \citep*{van19,ada15,ada14}.  In a half
page journal entry, \citet*{van17} noted that his accidentally 
discovered faint star with large proper motion had a spectral type
of ``about F0'' \citep*{van17}.  It was not until forty years(!) 
later that it became clear that van Maanen 2 was metal-poor with
respect to the Sun \citep*{wei60}.  Over the course of the another 
decade and a half, it became gradually clear that white dwarfs in 
general -- both hydrogen atmosphere degenerates void of metallic
features and helium atmosphere degenerates with photospheric metal
lines -- had heavy element abundances a few to several orders
of magnitude below solar \citep*{weh75,gre74,shi72,weg72,bue70}.  
Oxymoronically, white dwarfs with detectable, yet evidently 
sub-solar, heavy element abundances are now referred to as 
metal-rich.

The two fundamental classes of metal-poor-yet-rich white 
dwarfs are essentially the same as their metal-free counterparts,
which are designated by the main atmospheric constituent; either
hydrogen or helium.  It is easiest to think of white dwarfs as type
DA (hydrogen) or non-DA (helium).  For a wide range of temperatures,
white dwarfs with hydrogen atmospheres will manifest Balmer lines
(type DA), excepting perhaps the very hottest cases where hydrogen 
may be ionized (e.g. PG 1159 stars).  White dwarfs with helium 
atmospheres display a variety of spectral behaviors which depend
primarily on their effective temperatures; helium II lines (type 
DO) for the hottest, helium I lines (type DB) for a wide temperature
range, and no lines (type DC) for the coolest.  Although there do 
exist cases of mixed atmospheres, either inferred or directly observed,
typically one of the two light gases dominates the composition and 
photosphere; a notable exception is the case of carbon opacity in a
helium atmosphere (type DQ).  There are both subtleties and complexities
in the composition and spectral characteristics of white dwarfs which
will not be discussed here, but for the full guided tour of the white
dwarf spectral zoo, the reader is referred to both the definition of 
the current classification scheme \citep*{sio83} and a broad spectral 
atlas \citep*{wes93}.  The important points to take away from a given 
designated spectral type is: 1) the first letter after D indicates 
the dominant gas in the atmosphere; 2) when metals are detected in 
an ultraviolet or optical spectrum of a white dwarf, the letter Z 
is added to the designation (with DC becoming DZ).

\subsection{Metals Are Contaminants}

Any primordial heavy elements in white dwarfs can only be 
sustained in their photospheres for the brief ($t\sim10^7$ yr) 
period while the degenerate is still rather hot and contracting
significantly, and then only to a certain degree.  This is achieved 
through radiative levitation at $T_{\rm eff}>20,000$ K for hydrogen
atmospheres and $T_{\rm eff}>30,000$ K for helium atmospheres 
\citep*{cha95,fon79}.  Below these temperatures, the cooling 
degenerate stars develop significant convection zones, which 
enhance the gravitational settling of the elements, now unimpeded
by radiative forces \citep*{cha95,paq86,muc84,alc80,vau79,fon79,
fon76,sch58}.  Diffusion timescales for the sinking of metals are
always orders of magnitude shorter than the evolutionary (cooling)
timescales of white dwarfs.  Therefore, external sources are 
responsible for the presence of the metals within cool white 
dwarf photospheres.  

Although there were suspects and many spurious detections 
reported, all metal-rich white dwarfs were historically helium 
atmosphere degenerates until the confirmation of G74-7, the first 
DAZ star \citep*{lac83}.  Even so, a decade later all other known 
metal-bearing white dwarfs were still restricted to the helium-rich 
variety, despite reports to the contrary; an examination of Tables 
1 and 2 in \citet*{dup93b} with hindsight reveals that only G74-7 
remains classified as DAZ \citep*{zuc03,wol02,wes93}.  This situation 
began to change a little less than a decade ago with a couple of
individual discoveries, followed by many more, now totalling around 
fifty objects \citep*{koe05,zuc03,zuc98,hol97,koe97}.  The reason DAZ
stars are latecomers to the metal scene is because helium atmospheres
are quite transparent relative to hydrogen atmospheres; all else being
equal, a given calcium abundance will produce an equivalent width of
order $10^2-10^3$ times stronger in a helium as opposed to hydrogen 
atmosphere, making it far easier to infer the presence of photospheric
metals in helium-rich degenerates \citep*{zuc03,dup93b}.

Convection zones can be many orders of magnitude larger in helium
atmosphere white dwarfs, resulting in diffusion times for heavy 
elements up to $t\sim10^6$ yr \citep*{dup92,paq86,muc84,alc80,fon79,
vau79,fon76}.  This relatively long-lived photospheric retention
allows for the possibility that extant metals in such objects are
the remnant of a long ago (up to several diffusion timescales)
interstellar cloud encounter \citep*{dup93a,dup93b,dup92}. However,
it was shown that the Galactic positions and space motions of DZ
and DBZ stars are not correlated with local interstellar clouds
\citep*{aan93,aan85}.  Additionally with regards to cool helium 
atmosphere white dwarfs, there is the ever-present and still
unexplained lack of detectable hydrogen (or more specifically in
the case of those bearing metal lines, the very high inferred metal
to hydrogen abundance ratios), which should be expected in quantity,
and readily visible in low opacity helium-dominanted photospheres,
if accretion from the interstellar medium has occurred at any time
during their cooling \citep*{dup93b}.

The timecales for the diffusion of metals in hydrogen atmosphere 
white dwarfs diminish quite drastically relative to those in their
helium atmosphere counterparts, owing to significantly smaller 
convection zones for $T\ga6000$ K, and in warmer cases ($T\ga13000$ K)
where the convection zone is almost negligible, these diffusion times
can be just a few days \citep*{koe06,zuc03,paq86}.  Unless such objects
have just emerged from interstellar accretion episodes within these short
periods (unlikely), they must be currently accreting at rates sufficient
to produce the observed abundances.  \citet*{zuc03} find no correlation
between enhanced interstellar medium densities and the current positions
of accreting DAZ stars.  In contrast, \citet*{koe06} conclude that, with 
certain caveats, the warm, partially ionized medium can explain the 
observed accretion rates and abundances in DAZ stars.

\subsection{Minor Planets Are Metal-Rich}

Two decades ago it was proposed that accretion of circumstellar 
material might be the cause of the heavy metal abundances seen in
some white dwarfs, so that the origin of contaminating elements
seen is bimodal in nature; either interstellar or circumstellar 
\citep*{sio90,alc86}.  The first such model invoked episodic 
cometary impacts from reservoirs which managed to survive the
post-main sequence mass loss phases, particularly the asymptotic 
giant branch \citep*{deb02,par98,alc86}.  This particular model fails 
to explain 1) the DAZ stars with the highest metal abundances, and 2)
the observed distribution of abundances in general \citep*{zuc03}.
A more promising model of circumstellar accretion invokes the tidal
disruption of an asteroid, which goes on to form a ring of debris
around the white dwarf, from which the photospheric heavy elements
originate \citep*{jur03}.  Given that a typical solar system asteroid
is around $10^5$ times more massive than a typical comet, this model
can explain both the relatively high metal abundance and the observed 
infrared excess seen at several metal-rich white dwarfs \citep*{jur07a,
far07,bec05,jur03}.  While it may be the case that both mechanisms
create contaminated white dwarf photospheres -- perhaps circumstellar
in the case of high metal abundances and interstellar for those in 
the lower range -- there are a growing number of metal-rich white
dwarfs which are either confirmed or suspected to harbor 
circumstellar dust \citep*{jur07b,von07,kil06,jur06,rea05a}.  

Observations in the mid-infrared are most sensitive to both warm 
and cool orbiting dust at metal-bearing white dwarfs, providing a
direct test of circumstellar accretion hypotheses.  Additionally, 
such a search can constrain the frequency of orbiting material more 
strongly than ground-based near-infrared observations, which are 
sensitive only to dust not far from its sublimation temperature.

\section{OBSERVATIONS AND DATA}

\subsection{IRAC Imaging of DAZ Degenerates}

Table \ref{tbl1} lists the 17 metal-rich white dwarfs observed
with IRAC, taken from \citet*{zuc03} with the exception of GD 362
\citep*{gia04}.  All targets were chosen as DAZ stars, but recent
evidence implies that two of the white dwarfs (G77-50 and GD 362) are
helium-rich (\citealt*{koe05}; D. Koester 2007, private communication).  
For the purposes of this paper and the statistics which follow, many
targets are nominally referred to as DAZ, even though future observations
may reveal helium-rich atmospheres in some.  This fact is simply unavoidable
as helium becomes spectroscopically undetectable in white dwarfs cooler 
than $T_{\rm eff}\sim10,000$ K and its presence can only be inferred by
indirect methods. \citep*{ber92}.  These stars were chosen among available 
{\em Spitzer} Cycle 1 observations for their high metal abundances.

Between 2004 November and 2005 August, observations were executed
with the Infrared Array Camera (IRAC; \citealt*{faz04}) in all four
bandpasses: 3.6, 4.5, 5.7, and 7.9 $\mu$m.  A 20-point cycling dither
pattern (of medium step size) was used for each target in each bandpass,
with 30 s frame times at each position, yielding a total exposure time 
of 600 s at all wavelengths.  The data were processed with the IRAC 
calibration pipeline (versions $10-12$) to create a single, fully 
processed and reduced image ($1''.20$ pixels in all four channels) 
upon which to perform measurements.  Aperture photometry was carried
out with the standard IRAF task {\sf apphot}, and measured fluxes were
corrected for aperture size, but not for color.  Generally, the flux 
and signal-to-noise were measured in a $2-3$ pixel aperture radius, 
dependent upon target brightness and neighboring sources, with a $10-20$ 
pixel sky annulus.  The measured fluxes were converted to the standard 
IRAC aperture using corrections described in the most recent version of
the IRAC Data Handbook \citep*{ssc06}.  The results are listed in Table
\ref{tbl2}.

\subsection{Photometric Errors}

Listed in Table \ref{tbl2} are the total errors in the measured
and calibrated flux, together with IRAC pipeline versions with 
which the data were extracted from the archive.  The photometric
errors were estimated by taking the per pixel standard deviation 
in the extracted sky level and multiplying by the area of a 2 pixel
radius photometric aperture, the smallest radius for which there exist
derived aperture corrections in the IRAC Data Handbook \citep*{ssc06}.
This approach is conservative and does not assume Gaussian, random
noise for the following reasons.  It was found in general that the
observations went sufficiently deep as to approach or reach the 
confusion limit, primarily, but not exclusively, at the two shorter
wavelengths.  Additionally, there was sometimes low spatial frequency
structure seen in the background of the two longer wavelength images,
which appeared to be from either real diffuse sources such as cirrus,
or from imperfections in the IRAC pipeline.  These factors led to 
difficulty in the determination of the true sky level in the vicinity
of the science target, the per pixel noise in the sky, and in the flux
measurement itself.  Even in the smallest, 2 pixel radius, aperture 
used for photometry, there was occasional possible flux contamination
from neighboring sources in this relatively large 18.1 square arcsecond
area, compounded by the undersampled pixels.  This possibility was 
evident from a few (typical) to several (rare) percent increase in 
photometric flux that was sometimes seen as the aperture radius was 
increased from 2 to 3 pixels, which should not occur for single point 
sources and accurate aperture corrections.  All targets were 
unambiguously detected at all wavelengths.  

There are five other sources of photometric error in the IRAC
camera which were considered \citep*{ssc06}.  First, the absolute
calibration uncertainty in the IRAC instrument is reported as 3\%
\citep*{rea05b}.  Second, color corrections have been ignored and
are typically smaller than 1\%.  Third, there is the pixel phase 
dependent correction, which is reported to be no more than 4\% peak 
to peak or $\pm2$\% for a single image at 3.6 $\mu$m \citep*{ssc06}.  
The images analyzed here were produced by a set of 20 pseudo-random
dithers which essentially annihilates this source of error. Fourth,
there is the array location dependent correction for Rayleigh-Jeans 
type stellar sources.  This effect is the largest source of photometric
uncertainty in the IRAC instrument and may reach 10\% peak to peak or
$\pm5$\%, therefore a typical error of this type should be $2-3$\%.
However, experiments have shown that for well dithered data, as that
analyzed here, the effect tends to average out (as would be expected)
and is less than 1\% \citep*{car06,ssc06}.  

To assess the reliability of the absolute IRAC calibration, a couple of 
experiments were undertaken (M. Jura 2007, private communication).  First, 
the 3.6 to 4.5 $\mu$m flux ratios of the target stars with $T_{\rm eff}>7000$
K, and no evidence for the presence of warm dust, were examined and found to 
vary typically within 5\% of the mean, however a couple of targets deviated 
by as much as 10\% -- quite significant for such a small number of white 
dwarfs.  Second, a similar result was obtained by examining 4.5 to 7.9 $\mu$m
flux ratios of identically warm white dwarfs from \citet*{mul07}.  In this
instance the relatively lower signal-to-noise at 7.9 $\mu$m may have been a
factor as evidenced by the result of \citet*{tre07} who fit a subsample of
the \citet*{mul07} white dwarfs with current models to within 4\% at 4.5 
$\mu$m, yet only to within 10\% at 7.9 $\mu$m.  White dwarf models themselves
cannot account for the observed deviations in flux ratios at these warmer 
temperatures, which can only amount to around 1\% for extreme values of 
surface gravity and temperature (P. Bergeron 2007, private communication).
Furthermore, \citet*{hin06} report observed IRAC fluxes for 33 main sequence
stars which deviate from model predicted photospheric values, on average, by
8\% at 3.6 $\mu$m, 4\% at 4.5 $\mu$m, and 6\% at 7.9 $\mu$m.  \citet*{sil06}
find similar deviations between measured IRAC fluxes and model predictions 
for 74 young main sequence stars, amounting to 5\% on average at all 
wavelengths.  Based on these analyses and findings, it seems appropriate 
to assign a $1\sigma$ IRAC calibration uncertainty of 5\%.  For bright 
targets, the total error is dominated by the absolute calibration uncertainty, 
while for the faintest targets the total error is dominated by the uncertainty 
in the aperture photometry.  For these reasons, the total errors represented 
in Table \ref{tbl2} should be considered conservative.

\subsection{NIRI $L$-grism Spectroscopy of G29-38}

On three nights in 2006 January G29-38 was observed at Gemini
Observatory at Mauna Kea with the Near-Infrared Imager (NIRI; 
\citealt*{hod03}) in spectroscopy mode using the $L$-grism, 
which covers $3.0-4.1$ $\mu$m.  Spectra were taken at two positions 
along the $0''.75$ slit with 1 s exposures and 60 coadds.  Overall, 
approximately 2 hours of usable science frames were gathered.
Calibration frames taken each night included spectral flats as 
well as observations of one of the two A0V telluric standards HIP
110578 or HIP 12640, utilizing 0.2 s exposures and $10-20$ coadds
nodded at two positions along the slit.

Both the science and telluric standard frames from each night 
were pairwise subtracted at the two nod positions in order to 
best remove the bright, variable sky at these wavelengths.  The
subtracted frames were flat fielded and then median combined, with
bad pixels and cosmic rays fixed manually, creating two spectra 
of opposite polarity which were extracted using standard IRAF 
tasks.  The two extracted spectra from each night were wavelength
calibrated using aborption lines from the night sky, and averaged.
The science spectrum each night was divided by the telluric spectrum 
and multiplied by a blackbody of the appropriate temperature.  The 
science spectra from all three nights were shifted and averaged to 
create the final spectrum.  The flux was converted from $F_{\lambda}$
to $F_{\nu}$, normalized to one, then flux calibrated using IRAC
3.6 $\mu$m photometry.  The signal-to-noise was estimated by
measuring the standard deviation along 20 sections of 50 pixels
each spanning the entire wavelength range, yielding values between
4 and 10 (average 7) over $3.0-3.4$ $\mu$m, and between 8 and 15
(average of 12) over $3.4-4.1$ $\mu$m.  These were calculated
assuming continuum or pseudo-continuum over all regions.

\section{ANALYSIS AND RESULTS}

\subsection{Spectral Energy Distributions}

The IRAC fluxes for all targets, together with optical and
near-infrared data, are plotted as spectral energy distributions
in Figures \ref{fig1}--\ref{fig5}.  In general, the short wavelength
photometry was taken from the most accurate and reliable sources 
available, including but not limited to: \citet*{mcc03} and references
therein; 2MASS \citep*{skr06}; DENIS \citep*{den05}; and various other
sources \citep*{lie05,mon03,zuc03,ber01,ber97}.  Because the aim of the
survey was to identify significant mid-infrared photometric excess due 
to opaque dust, blackbody fits to the spectral energy distributions of 
the target white dwarfs will suffice to model their expected, essentially 
Rayleigh-Jeans, behavior at IRAC wavelengths.  If the total uncertainty
in the IRAC photometry decreases sufficiently, white dwarf models should 
prove more useful in this regard.

Of the 17 observed stars, three display excess radiation within their 
IRAC beams (considered here to be the solid angle contained within a
full width at half maximum Airy disk, typically $2''.0-2''.4$ in diameter and 
$3.1-4.5$ square arcseconds in area at $3-8$ $\mu$m), with a high degree of
certainty.  The case of G166-58 is discussed in some detail below and shows
evidence for continuum emission from $T\sim400$ K dust.  Both G29-38 and 
GD 362 show warm ($T\approx900$ K) thermal continuum and strong 
silicate emission in their mid-infrared spectra and have therefore been 
confirmed to harbor orbiting rings of dust \citep*{jur07a,rea05a}.  For the
remainder of the targets, the IRAC observations either rule out emission from 
warm, opaque debris or are of sufficiently low signal-to-noise as to preclude
a definitive conclusion either way.  Unfortunately, there was a single target
(G21-16) for which accurate fluxes could not be extracted due to unavoidable
source confusion within and around the IRAC beam.  An unsuccessful attempt 
was made to photometrically isolate the white dwarf with the IRAF task 
{\sf daophot}; based on the signifiant crowding, it is likely that the 
IRAC beam itself is contaminated at all wavelengths.

From the present work, 3 of 17 metal-rich white dwarfs, or 18\% display 
definite IRAC continuum excess consistent with orbiting dust.  However,
this fraction is only 2 in 15 DAZ stars, or 13\%.  If one counts all DAZ
white dwarfs observed with IRAC in Cycle 1, the total which display 
IRAC flux excess is 3 of 25 targets or 12 \% \citep*{von07,kil06}.

\subsection{GD 362}

The spectral energy distribution of GD 362, together with its 
IRAC data, is displayed in Figure \ref{fig5}.  As expected on
the basis of ground-based $JHKL'N'$ photometry \citep*{bec05},
the $3-8$ $\mu$m region is dominated by warm thermal emission from
orbiting dust, whose flux can be reproduced out to $\lambda\approx6$
$\mu$m by a $T=900$ K blackbody.  This approximation is physically
nonviable as it corresponds to a ring (or sphere) with a single
temperature and radius, yet does not deviate much from a ring 
model of substantial radial extent \citet*{jur03} at these shorter 
mid-infrared wavelengths.  The measured flux at 7.9 $\mu$m has not 
been corrected for color and shows an indication that the broad and 
strong 10 $\mu$m silicate emission feature \citep*{jur07a,far07} 
contributes significantly into that relatively wide ($\Delta
\lambda=6.5-9.5$ $\mu$m) bandpass.  

In Figure \ref{fig6}, the $2-24$ $\mu$m photometric data on 
GD 362 are plotted together with the more realistic model of 
\citet*{jur03}.  The model invokes a face-on geometrically thin 
opaque dust ring of finite radial extent, with an inner temperature 
of $T_{\rm in}=1200$ K and an outer temperature range $T_{\rm out}=
300-600$ K.  For GD 362, these temperatures correspond to a ring 
which extends from $D_{\rm in}\approx0.1$ $R_{\odot}$ to $D_{\rm
out}\approx0.3-0.7$ $R_{\odot}$ \citep*{chi97}.  The data agree 
reasonably well with the higher temperature curve, excepting the
24 $\mu$m flux (disregarding the silicate emission-enhanced 7.9 
$\mu$m flux).  There are two possibilities based on this model: 
1) the outermost opaque grains have a temperature $T_{\rm out}<
600$ K or; 2) the flux at 24 $\mu$m is affected by another dust 
emission feature (which would indicate forsterite, if present).  
Photometric and spectroscopic observations of GD 362 utilizing
all three instruments aboard {\em Spitzer}, including a detailed
model fit of the orbiting dust, its mid-infrared thermal continuum
and emission features, dust mass estimates, temperature and particle
size distribution, dimensions and types of the emitting regions 
are presented in \citet*{jur07a}.

\subsection{G29-38}

The spectral energy distribution of G29-38, together with its 
IRAC data, is displayed in Figure \ref{fig5}.  While the overall 
similarity between G29-38 and GD 362 is apparent, there are some
distinctions.  Although their inner dust temperatures are clearly
similar, the thermal continuum flux of G29-38 over $3-6$ $\mu$m
appears to be falling, while for GD 362 it is rising; likely an
indication of varying amounts of opaque dust both slightly warmer
and cooler than the $T=900$ K blackbody approximations for the
excess at each of these stars.  In order to fit the slope of the
$3-6$ $\mu$m photometry of G29-38 with a blackbody, a temperature 
near 1100 K is necessary, greatly overpredicting the near-infrared
flux.  Yet as can be seen from the figure, the 3.6 $\mu$m flux is
somewhat underpredicted by 900 K. 

Using the more plausible and physical model of \citet*{jur03},
Figure \ref{fig6} plots the thermal infrared excess of G29-38 
from $2-24$ $\mu$m from all available {\em Spitzer} photometric
data, together with the model curves.  These disk models are
exactly those applied to G29-38 in \citet*{jur03}, now plotted
with more accurate data with greater wavelength coverage.  For
the same inner and outer temperatures given above, the extent of
the opaque ring is from $D_{\rm in}\approx0.1$ $R_{\odot}$ to 
$D_{\rm out} \approx0.4-0.9$ $R_{\odot}$ for G29-38.  In this
case, unlike GD 362, all the photometric data are fitted decently
by the model where the outermost grains have a temperature near 
600 K, with the possible exception of the 3.6 $\mu$m flux.

The apparently excessive flux from G29-38 at the shortest 
wavelength IRAC channel was first noticed when the data were 
initially retrieved from the {\em Spitzer} archive using pipeline
version 11.0.  It was present again one version later, and then 
triple checked in 2006 November with version 14.0, which was used 
for all the IRAC data on G29-39 in this work.  The deviation 
between the models and the measured flux at 3.6 $\mu$m is $0.92-
0.98$ mJy from the best fit ring model or the blackbody.  Taking 
photometric error into account, which is entirely due to instrument 
calibration uncertainty, the average deviation is 0.70 mJy; 
approximately $2.8\sigma$ of the flux error, or 10\% of the excess.
$3-4$ $\mu$m spectroscopy was undertaken to investigate possible
sources of this extra emission, and to better assess if it is real.

As can be seen from Figure \ref{fig7}, the $L$-grism spectrum
of G29-38 is essentially featureless, with an apparent, slight,
upward slope towards 4 $\mu$m (a thermal continuum approximated by
a 900 K blackbody would peak longward of 5 $\mu$m in $F_{\nu}$).
Previous to this investigation, the only spectral information in 
this wavelength regime came from data presented in \citet*{tok90}, 
where a relatively low signal-to-noise spectrum is not inconsistent
with emission over $3.2-3.7$ $\mu$m.  The spectral flux error in
the Figure \ref{fig7} data, translated to mJy via the IRAC 3.6 
$\mu$m flux, is typically 1.2 mJy over $3.0-3.4$ $\mu$m (outside
the atmospheric transmission window, an area very sensitive to water
vapor and prone to large variability), and 0.7 mJy over $3.4-4.1$ 
$\mu$m (atop the $L'$-band).  

There exist a plethora of emission features in this region
associated with polycyclic aromatic hydrocarbons and their 
multitudinous close relatives: specifically, features between 
3.2 and 3.6 $\mu$m seen in the interstellar medium \citep*{dra03,
all89,geb85}; circumstellar matter \citep*{mal98,bei96,geb85};
ultraviolet-excited (planetary, proto-planetary, and reflection)
nebulae \citep*{geb92,geb89,geb85}; as well as in comets 
\citep*{boc95,bro91,baa86}.  The well known 3.3 $\mu$m feature,
when present, is always significantly weaker than its other family
members at 6.2, 7.7, 8.6, and 11.3 $\mu$m \citep*{dra03,mal98,bei96,
all89}.  Because these stronger features are absent from the IRS
$5-15$ $\mu$m spectrum of G29-38 \citep*{rea05a}, it is expected
no feature should be present at 3.3 $\mu$m, although the IRS data
were published after the the $L$-grism observations were planned.
There are additional features around 3.4 $\mu$m (primarily due to
methanol, ethane, and other hydrocarbon species) which typically 
dominate this region when observed towards comets \citep*{mum01,
cro97,boc95,baa86}, but which are much weaker than the 3.3 $\mu$m
feature in circumstellar environments \citep*{mal98,bei96}.  Because
of these observational facts, in addition to the relatively fragile 
nature of hydrocarbons (some small species are more volatile than
water ice) in the vicinity of high density ultraviolet radiation 
fields \citep*{job97,bro91}, it is unlikely that a cometary feature 
would be seen in the vicinity of the $0.1-0.4$ $R_{\odot}$ 
circumstellar dust ring at G29-38.

The failure to detect any possible sources of excess emission 
in the $3-4$ $\mu$m region at G29-38 leaves a few possibilities
for the 2.8 $\sigma$ disagreement between its measured IRAC flux 
and the applied models.  The first is that the photometry in the 
3.6 $\mu$m bandpass is improperly calibrated, making it inaccurate.
The second is that both the ring model of \citet*{jur03} and the 
single temperature blackbody fail to predict the correct flux at 
this wavelength.  A $T=1000$ K blackbody (Figure \ref{fig6}) is
not inconsistent with the $2-6$ $\mu$m photometry, but underpredicts
the three longer wavelength {\em Spitzer} data points, requiring that
dust emission affects those bandpasses.  The third possibility is the
discrepancy arises from photometric variability.  It is well-known 
that G29-38 is a pulsating white dwarf; $B$-band light curves reveal
periods of 615, 268, 243, and 186 s with amplitudes of 0.12, 0.03, 
0.03, and 0.02 mag respectively, which are mirrored at $K$-band with
corresponding strengths of 0.02, 0.02, 0.03, and 0.03 mag respectively 
\citep*{pat91,gra90}.  Although matching periods were searched for at
$L$, none were found, but with sufficiently large upper limits which
do not exclude variations similar to those seen at $K$ \citep*{pat91}.
Because the IRAC observations lasted 600 s, these photometric variability
timescales cannot explain the 3.6 $\mu$m flux being possibly high.
Given that the IRAC channel 1 ($\Delta \lambda=3.2-3.9$ $\mu$m; 
\citealt*{faz04}) data agree reasonably well with ground-based 
$L$-band ($\Delta \lambda=3.2-3.8$ $\mu$m; \citealt*{tok00}) data
with no color corrections to either, the discrepancy is likely to
be model disagreement.  The 4.5 and 7.9 $\mu$m fluxes for G29-38
in Table \ref{tbl2} agree well with the quoted (uncorrected for 
color) fluxes reported by \citet*{rea05a}.  It is possible that
the deviation arises from variability which has not yet been seen
or reported, but no such claim is being made based on the present
data.  A fourth possibility might be a companion, but all substellar
models of appropriate age range predict about twice as much excess
at 4.5 $\mu$m relative to 3.6 $\mu$m, as well as a commensurate rise
from 3 to 4 $\mu$m which is not observed in the $L$-grism spectrum
\citep*{bar03,bur03}.

\subsection{G166-58}

Figure \ref{fig4} displays the spectral energy distribution of
G166-58 together with its IRAC flux measurements.  The plotted data
are optical $BVRI$ from \citet*{ber01}, the average of 2 $U$-band 
values cited in \citet*{mcc03}, and near-infrared $JHK$ which are
the average of values given in \citet*{zuc03} and \citet*{ber01}.
The effective temperature of the white dwarf is taken from both
\citet*{lie05} and \citet*{zuc03}, where the determinations are
both within 0.6\% of 7400 K.  G166-58 displays clear excess emission
within its IRAC beam beginning at 5 $\mu$m, making it unique among
white dwarfs confirmed or suspected to harbor orbiting dust.  In 
all other cases, the excess becomes unambiguous by 3 $\mu$m, as in
G29-38 and GD 362 \citep*{von07,kil06,bec05,zuc87b}.  The implied 
temperature of the $5-8$ $\mu$m excess can be reproduced by a blackbody 
of $T\approx400$ K, and is substantially cooler than the $T\approx900$
K temperatures inferred in all other dusty white dwarfs with $2-3$
$\mu$m excess.  Yet circumstellar dust at 400 K is still considered 
warm relative to the overwhelming majority of main sequence stars
with infrared excess attributed to a debris disk; with a few notable
exceptions, these are all Kuiper belt analogs with $T_{\rm dust}\la
120$ K, and $D_{\rm dust}\ga10$ AU \citep*{bei06,son05,lau02,zuc01,
che01}.

Before proceeding further with any analysis and interpretation,
owing both to the uniqueness of the IRAC data on G166-58 and the
IRAC field in its vicinity, the nature and validity of the excess 
must be examined.

\subsubsection{IRAC Beam Contamination}

There is another source $5''.3$ from G166-58 seen in all four IRAC
images, which are displayed in Figure \ref{fig8}.  This source is also 
present in the SDSS Photometric Catalog (release 5) where it is 
designated SDSS J145806.96+293726.3 and classified as a galaxy 
\citep*{ade07}.  This object has optical and mid-infrared colors that 
are consistent with an extragalactic source.  The IRAC images of G166-58 
and the nearby galaxy overlap near $2''.5-3''.0$ from their image centers.  
Both sources appear point-like at all four wavelengths.  The percentage of 
flux from the nearby galaxy in a 2 pixel radius centered at G166-58, measured 
relative to the total flux of the white dwarf over the same area, is relatively benign 
at $2-3$\%, with the exception of 7.9 $\mu$m where it is 12\%.  This was determined 
by folding the IRAC point spread function of G166-58 in two along the axis of 
symmetry (up-down in the Figure \ref{fig8} images), and subtracting one 
side from the other.  Exploiting this symmetry, it was straightforward to remove 
the contamination in the aperture photometry of G166-58 and vice versa
for similar measurements of the nearby galaxy (its flux is 0.18, 0.17, 
0.13, and 0.43 mJy at 3.6, 4.5, 5.7, and 7.9 $\mu$m respectively).  
The values in Table \ref{tbl2} for G166-58 correspond to the signal 
after removal of this unwanted flux.

Two additional methods used to obtain the flux of G166-58 are discussed 
below, and a comparison of the fluxes obtained for the white dwarf by each 
method is shown in Table \ref{tbl3}.  First, the white dwarf and nearby galaxy 
were photometrically fitted and spatially deconvolved using {\sf daophot}.
Second, photometry was obtained by radial profile analysis of the white dwarf.  
Figure \ref{fig9} displays overlaying and identical linear contour plots for 7.9 $\mu$m 
images of G166-58 both before and after the removal of the nearby galaxy.  The 
parameters from a 2.0 pixel gaussian radial profile fit (image centroid and full 
width at half maximum) of the white dwarf remain essentially unchanged by 
inclusion of the galaxy.  Radial profile analysis of G166-58 at each IRAC channel 
yields a measurement for the white dwarf, via comparison with the radial profiles 
and fluxes of G29-38 and GD 362.  Examination of G166-58 in a 7.9 $\mu$m image 
where only the galaxy is fitted and subtracted reveals a clear point-like source 
and excess at the location of the white dwarf.  In summary, Table \ref{tbl3} illustrates
excellent agreement among the three photometric methods used to determine the
flux of G166-58, with the exception of {\sf daophot} at 3.6 $\mu$m (which may be
due to undersampled data, which is most germaine at this shortest wavelength.)

Further support that the two IRAC sources are separated photometrically 
is evidenced by their 5.7 to 7.9 $\mu$m colors; the nearby galaxy has a flux 
ratio of 0.30, while the excess detected at G166-58 has a flux ratio of 0.46.  
Hence, with regards to the adjacent galaxy, there should be no doubt of its 
lack of influence in the white dwarf data presented. 

The known density of background galaxies, along with the presence of 
the relatively bright nearby source on the IRAC chips, prompted an evaluation 
of the probability that yet another, hidden source could be contaminating flux 
measurements within the beam of G166-58 itself, and therefore be responsible 
for the $5-8$ $\mu$m excess.  The white dwarf is located at galactic latitude 
$b=+68\arcdeg$, hence any contaminating source would almost certainly be 
extragalactic in nature.  To assess the probability of such a line of sight coincidence, 
the following estimations were made.  The 7.9 $\mu$m excess emission at G166-58 
is about 0.07 mJy, or 14.9 mag.  {\em Spitzer} IRAC 7.9 $\mu$m source counts from 
\citet*{faz04b} indicate approximately 3,000 galaxies per magnitude per square 
degree at $15^{th}$ magnitude.  Taking the distribution to be flat over a 0.5 magnitude 
interval centered at 15.0 mag, this yields around 1500 galaxies per square degree 
of appropriate brightness to reproduce the 7.9 $\mu$m excess seen at G166-58.  
Because the white dwarf and its excess display a point-like nature at this wavelength
(see Figure \ref{fig9}), any unresolved background source would have to lie within
a small fraction of the IRAC beam width of G166-58 in the plane of the sky,
certainly within an area of 2 square arcseconds, conservatively speaking.
Therefore, the probability of finding a galaxy of 0.07 mJy brightness at 
7.9 $\mu$m within 2 square arcseconds of G166-58 is about 1 in 4300.  The
odds that 1 in 17 target stars is contaminated by such a background galaxy
should then be around 1 in 250.

Therefore, the most likely explanation for the excess $5-8$ $\mu$m emission 
at G166-58 is circumstellar dust associated with the white dwarf.

\subsubsection{Circumstellar Dust}

If the $5-8$ $\mu$m IRAC fluxes are the sum of the white dwarf
and another source, one can immediately rule out a cold companion.
A substellar object with an energy distribution similar to that 
implied by the 400 K blackbody fit, would have a radius of roughly 
$2\times10^8$ m, more than twice the size of Jupiter.  Furthermore,
the combined spectral energy distribution appears nothing like 
what might be expected from such an orbiting cold degenerate; 
it most notably lacks significant flux expected at 4.5 $\mu$m
\citep*{far05b,bur03}.  Therefore, the excess emission must 
be due to warm circumstellar material.

As a first stab at modeling the excess, one might invoke tidal
dust rings similar to those which have been successful for both
G29-38 and GD 362 \citep*{jur07a,jur03}.  Following the formalism
of \citet*{chi97}, a flat opaque ring (or disk) geometry, implies
a dust grain temperature-radius relation given by

\begin{equation}
T_{\rm gr} \approx { \left( \frac{2}{3 \pi} \right) }^{1/4}
	   	   { \left( \frac{R}{D}     \right) }^{3/4} T_{\rm eff}
\end{equation}

\noindent
where $T_{\rm gr}$, $D$, $T_{\rm eff}$, and $R$ are the temperature
of the emitting grains, their distance from the star, the stellar 
effective temperature, and the stellar radius, respectively.  If one
assumes that G166-58 is a single, carbon-oxygen core white dwarf with
log $g=7.97$ and $T_{\rm eff}=7390$ K \citep*{lie05}, then its radius
is $R=0.0132$ $R_{\odot}$ \citep*{ber95a}.  These stellar parameters
yield a distance of $D_{\rm in}=0.38$ $R_{\odot}$ to 400 K grains.
If one assumes the ring extends to where the grain temperature is
200 K, then the ring would extend to $D_{\rm out}=0.97$ $R_{\odot}$.
These inner and outer disk radii would be significantly larger than
those implied for G29-38, where the ring probably does not extend 
much further than $D_{\rm out}\approx0.4$ $R_{\odot}$ (around 30
stellar radii) for outer grain temperatures of $T_{\rm out}=600$ K,
which seem to fit the data in Figure \ref{fig6} quite well.  This 
is also true of the implied size of the tidal ring about GD 362, 
where models \citep*{jur07a} also yield an outer radius $D_{\rm 
out}\approx0.4$ $R_{\odot}$ (around 40 stellar radii).

Regarding the outer edge of the disk and an appropriate scale for
the tidal breakup of a rocky body such as an asteroid or comet, 
\citet*{dav99} provides a thorough review and revision of effective
Roche limits.  The distance, $\delta$, at which a small orbiting 
body will be disrupted by the gravitational field of a large body
of radius $R$, can be expressed, in simplified form, as

\begin{equation}
\delta \approx \alpha { \left( \frac{P}{\rho} \right) }^{1/3} R
\end{equation}

\noindent
where $P$ and $\rho$ are the densities of the large and small
bodies, and $\alpha$ is a constant which typically has a value
in the range $1-2.45$, but can be smaller.  The coefficient $\alpha$ 
depends on the model, which may include factors such as; composition,
heterogeneity, size, shape, rotation, orbital characterisitcs, 
shear and tensile strengths.  The classical value of $\alpha=2.45$ 
is for the case of a rotating, uniform, self-gravitating liquid in
a circular orbit, whereas for a nonrotating, spherical satellite 
of solid rock or ice, the value becomes $\alpha=1.26$.  This last 
approximation is likely to be valid for typical asteroids and comets
with radii $r\ga5$ km, at least to within 50\% \citep*{dav99,bos91}.
Taking $\rho=1$ g ${\rm cm}^{-3}$, which lies within the range of
densities for both asteroids and comets ($\rho_{\rm ast}=1.0-3.5$
g ${\rm cm}^{-3}$, $\rho_{\rm com}=0.1-1.1$ g ${\rm cm}^{-3}$;
\citealt*{bin00}), and calculating the average density of G166-58 
with the parameters above, an estimate of the Roche limit using 
$\alpha=1.26$ is $\delta\approx1.2$ $R_{\odot}$.  Hence, if an
optically thick, flat ring orbits G166-58 at distances corresponding
to 200 K dust grains, these would lie within a region consistent with
tidal disruption of a minor planet.  However, at grain temperatures
of 100 K, the implied distance from G166-58 would be 2.4 $R_{\odot}$,
a region where asteroids or comets should remain intact, even if
liquified \citep*{roc48}.

Integrating the flux of the 400 K blackbody fit to the IRAC excess
yields $L_{\rm IR}=1.0\times10^{-6}$ $L_{\odot}$, while the stellar
luminosity given by models for a hydrogen atmosphere white dwarf 
with $T_{\rm eff}=7400$ K and log $g=8$, yields $L=4.4\times10^{-4}$ 
$L_{\odot}$ \citep*{ber95a}.  Together, these determine $\tau=L_{\rm 
IR}/L=0.0023$, which is logarithmically about midway between $\tau
\approx0.0002$ found for the main sequence A3 star $\zeta$ Leporis, 
and $\tau\approx0.03$ for both G29-38 and GD 362 \citep*{jur07a,bec05,
rea05a,che01}.  Therefore, instead of invoking an optically thick, 
flat ring, it may be more appropriate to suppose an optically thin
shell of blackbody grains in radiative equilibrium, whose distance
is given by \citep*{che01}

\begin{equation}
T_{\rm gr} \approx { \left( \frac{R}{2D} \right) }^{1/2} T_{\rm eff}
\end{equation}

\noindent
In this case, 400 K grains would be located near 2.3 $R_{\odot}$,
about 6 times further out than predicted by the opaque disk model
and well beyond the Roche limit for large rocks.  An advantage of 
this assumption is that it allows an estimation of the minimum dust 
mass contained in the disk.  Because white dwarfs have masses $M\sim1$ 
$M_{\odot}$, yet greatly reduced luminosities $L\sim10^{-2}-10^{-4}$ 
$L_{\odot}$, radiation pressure on dust grains cannot compete with 
gravitational attraction.  This can be seen by examining the ratio 
of these two forces, represented by the parameter

\begin{equation}
\beta = \frac{3}{16\pi} \frac{LQ_{\rm pr}}{GMc\rho a}
\end{equation}

\noindent
where $a$ is the dust particle radius and $Q_{\rm pr}$ is the 
radiation pressure coupling coefficient \citep*{art88}.  Assuming 
the case of geometric optics, where the effective grain cross section 
equals its geometric cross section and $Q_{\rm pr}\approx1$, yields the
maximum possible value for $\beta$ near $a=0.1$ $\mu$m.  For $\rho=1$ g 
${\rm cm}^{-3}$, $L=4.3\times10^{-4}$ $L_{\odot}$, $M=0.58$ $M_{\odot}$, 
$\beta_{\rm max}=0.004$ and hence gravitational forces dominate over 
radiation pressure.  This simply serves to show that sub-micron size 
dust and gas particles could certainly be present at white dwarfs 
without any danger of being lost to radiation pressure.  The minimum
dust mass of an optically thin disk is approximately

\begin{equation}
M_{\rm dust} \approx \frac{16\pi}{3} \frac{L_{\rm IR}}{L} D^2 \rho a
\end{equation}

\noindent
Although both smaller and larger particles are almost certainly present,
the size of the thermally emitting dust is on the order of $1-10$ $\mu$m.
Taking the density of silicate grains to be 2.5 g ${\rm cm}^{-3}$ gives a
rough lower limit to the mass of the the dust disk of $M_{\rm dust}\sim2
\times10^{18}$ g, a mass of a very large comet.  While radiation pressure
cannot remove dust at white dwarfs, drag forces can.  The timescale for 
Poynting-Robertson removal of particles is given by \citep*{bur79}

\begin{equation}
t_{\rm pr} = \frac{4\pi}{3} \frac{c^2 D^2 \rho a}{LQ_{\rm pr}}
\end{equation}

\noindent
For white dwarfs in general, the ratio $D^2/L$ for optically thin dust
will be $10^2-10^3$ times smaller than for main sequence stars.  Silicates
of 1 $\mu$m size and $\rho=2.5$ g ${\rm cm}^{-3}$ orbiting G166-58 at 2.3 
$R_{\odot}$ will be removed by the drag force in $t_{\rm pr}=460$ yr.  The
ratio $Q_{\rm pr}/a$ increases to a maximum near $a=0.1$ $\mu$m, implying 
removal timescales about 10 times more rapid.  For smaller particles, the 
radiation coupling efficiency swiftly declines and this ratio decreases, 
levelling off near $a=0.01$ $\mu$m at a value similar to that for 1 $\mu$m 
grains \citep*{art88}.  Hence, in an optically thin disk, all particles up 
to 1 $\mu$m in size should be removed by Poynting-Robertson drag within 500
yr.  Assuming a present balance between accretion and diffusion with the 
above stellar parameters for G166-58, the rate at which it currently gains
refractory circumstellar material is $\dot{M}=2.0\times10^{8}$ g ${\rm s}^{-1}$ 
(Koester \& Wilkens 2006; where a factor of 0.01 has been included to reflect 
the absence of accreted hydrogen and helium).  If this accretion occurs over
a single diffusion timescale of $10^{3.1}$ yr, the total accreted mass would
$8\times10^{18}$ g, equivalent to the mass of a small solar system asteroid. 

\subsubsection{Double Degeneracy}

Based on available data, it appears possible or perhaps likely
that G166-58 is not a single white dwarf.  Although this target was
not listed in Table 2 of \citet*{zuc03} for white dwarfs known or 
suspected to be in binary systems, they determine a low surface 
gravity of log $g=7.58$ via a combination of optical through 
near-infrared photometry and parallax.  At 7400 K, contemporary 
white dwarf models predict that such a low surface gravity implies
a radius $R\approx0.016$ $R_{\odot}$, which is about 25\% too large
for a normal carbon-oxygen core degenerate \citep*{ber95a}.  An
inferred oversized radius and corresponding overluminosity can be
explained either by a single, low mass, helium core degenerate or 
via binarity involving two components of similar ($\Delta m\la1$ mag) 
brightness -- a double degenerate.  Actually, the fact that G166-58 
appears overluminous based on its spectral energy distribution and 
parallax has existed since the analysis of \citet*{ber01}.  There,
using an essentially identical procedure, a low surface gravity of
log $g=7.66$ was determined, carrying the same implication.  The
near-infrared $JHK$ measurements of both \citet*{zuc03} and
\citet*{ber01} agree to within their respective errors.

Because the implied overluminosity of G166-58 relies fairly heavily
on its trigonometric parallax(es), a careful literature search was
performed in order to assess all available astrometric data.  The
value employed by \citet*{ber01} comes from the most recent version
of the Yale Parallax Catalog, and is given as $\pi=0''.0289\pm0''.0041$
\citep*{van95}.  The source of the Yale catalog parallax is the US Naval
Observatory (H. C. Harris 2006, private communication), first published 
in \citet*{rou72} and then updated (and possibly revised) in \citet*{har80}.
This parallax is given as $\pi_{\rm relative}=0''.0277\pm0''.0041$ and
$\pi_{\rm absolute}=0''.0298$ in their Table 1.  \citet*{mcc03} quote a
value of $\pi=0''.028$ while citing \citet*{rou72}, but this is merely
the {\em relative} (i.e. measured) parallax quoted above.  Paying careful
attention to detail, one can see the Yale and US Naval Observatory values
are not identical, but reflect slightly different astrometric corrections 
to obtain the absolute parallaxes, with the Yale galactic model being more
recent and likely more reliable.  Therefore, the only existing parallax 
measurement for G166-58, upon which its overluminosity rests, is $\pi_{\rm
trig}=0''.0289\pm0''.0041$ or $d=34.6^{+5.7}_{-4.3}$ pc.

\citet*{ber01} use this parallax for G166-58, together with its 
optical and near-infrared spectral energy distribution to deduce
$T_{\rm eff}=7310$K, $M=0.41$ $M_{\odot}$, and $M_V=12.90$ mag.  
However, fitting models to the slope and Balmer line profiles of 
an optical spectrum, \citet*{lie05} determine a nearly identical
temperature but log $g=7.97$, $M=0.58$ $M_{\odot}$, $M_V=13.28$, 
and $d=29.1$ pc instead.  This disparity still can be explained by
the presence of another white dwarf which contributes the extra flux
at $V$, but {\em not} by a single low mass white dwarf with an overly
large radius, which is inconsistent with the Balmer line profile fit.
The apparent discrepancy might also be explained by supposing that
G166-58 is located closer to 29 pc, which is near the 30.3 pc lower 
limit implied by the uncertainty in its parallax.  \citet*{zuc03} 
remark that radial velocity measurements in 1998 June, 1999 April 
and July agree within their errors and hence there is some weak 
evidence against radial velocity variability in G166-58.  A DA or 
DC white dwarf companion at $\Delta V\approx1$ mag would dilute the
Balmer lines and possibly redden the optical spectrum, depending on
its effective temperature \citep*{ber90}.  It may be the case that
such a companion should have already caused notice in the data 
analyzed by \citet*{lie05}, but nothing was noted.

Figure \ref{fig4} appears to imply that a single temperature
blackbody does not fit the optical and near-infrared data perfectly.
In fact, there seems to be a slight near-infrared excess at $JHK$,
which would become more prominent if a higher temperature model 
were used to fit $UBVRI$ only.  Yet the IRAC $3-4$ $\mu$m data lie 
very close to the plotted blackbody, so perhaps there is another 
explanation for any apparent mismatch between model and published
flux, such as data which are not photometric or calibration errors.
In any case, follow up observations of G166-58 would help to evaluate
the possibility that it may be a double degenerate.  Specifically, 
a radial velocity study, a careful model analysis of its spectral
energy distribution with accurate and precise photometry, or
another trigonometric parallax measurement would all be useful.

\subsubsection{Circumbinary Debris}

Identification of a double degenerate suspect follows more or 
less as it does for main sequence stars; the object lies superior
to its expected position in a Hertzsprung-Russell (or equivalent)
diagram.  This requires that the distance and effective temperature
of the star are known or constrained in some fashion.  For white 
dwarfs in the field, the nominal sequence is located near and about
a radius corresponding to $M=0.6$ $M_{\odot}$, or log $g=8$ for cool
to warm white dwarfs \citep*{ber01,ber95c,ber92}.  A single target
which lies above the log $g=8$ sequence can either be a single white
dwarf responsible for the bulk of the luminosity via an overly large
radius (and a comparably low mass and surface gravity), or a near
equal brightness binary.  Somewhat oxymoronically, single, low mass
($M<0.45$ $M_{\odot}$, helium core) white dwarfs are understood to be
the end products of close binary evolution and more likely than not 
still attached to their stellar cannibals \citep*{han98,mar95,ber92}.  
But a single white dwarf with low surface gravity will have appropriately
thinned Balmer lines due to a reduction in Stark broadening, and this is
not seen in G166-58 \citep*{lie05}.  Therefore, if G166-58 is binary, it
must be composed of two relatively normal mass white dwarfs of similar
brightness.

Given the fact that several such double degenerate suspects have
turned out to be bona fide \citep*{far05a,zuc03,ber01,mar95}, it is
appropriate to consider dust models which conform to this distinct
possibility in addition to those above for a single white dwarf (see
Table \ref{tbl4} for a summary of possible parameters).  The average 
stellar parameters for G166-58 from the two analyses which utilize its
trigonometric parallax, both of which find a similar overluminosity, 
are log $g=7.62$, and $T_{\rm eff}=7340$ K \citep*{zuc03,ber01}.  The 
resulting effective radiating surface for such a solution is $R=0.0163$
$R_{\odot}$ \citep*{ber95a}.  Applying the opaque, flat ring model and
Equation (1) with these stellar parameters yields a distance of $D_{\rm
in}=0.47$ $R_{\odot}$ to 400 K grains.  This is a valid model for a
single, low mass white dwarf with the above large radius, but not for
a binary.  In short, it is not possible to fit two similar white dwarfs
within such a tight disk while avoiding interactions (mass transfer) and
at the same time maintaining relatively cool, and sufficiently distant
400 K grains.  Additionally, there would be gravitational interactions 
between the binary and such a disk which would probably preclude a flat 
geometry.  Returning to the optically thin case, Equation (3) places 400 
K grains at 2.7 $R_{\odot}$.  With some adjustment of parameters, this 
scenario allows ample space for a double degenerate to orbit without 
interaction \citep*{morr05}, and sufficient distance from the binary 
in order to maintain 400 K dust.

Observations of G166-58 with the Fine Guidance Sensors aboard the
{\em Hubble Space Telescope} show that it is spatially unresolved 
to approximately $0''.008$ (E. P. Nelan 2006, private communication).
This precludes separations wider than 0.3 AU or 60 $R_{\odot}$, yet
still permits binarity with dust located at one white dwarf component.

\subsection{Metal-Rich Double Degenerates}

In Table \ref{tbl1}, there are three confirmed or suspected double 
degenerates: G77-50, EC 1124$-$293, G166-58.  These white dwarfs are 
relisted in Table \ref{tbl5} together with their divergent spectroscopic 
and photometric parameters.  In light of the potential binarity of 
G166-58, a brief focus on the other similar systems is appropriate.

{\em G77-50}.  This white dwarf has a recent parallax measurement 
from \citet*{sma03} of $\pi_{\rm trig}=0''.0595\pm0''.0032$ over 
a 6.2 yr baseline, placing it firmly at $d=16.8^{+1.0}_{-0.9}$ pc. 
Thus, its derived spectroscopic parameters in Table \ref{tbl5}, 
which consist of a good $T_{\rm eff}$ determination from optical 
and near-infrared photometry, but only a crude log $g$ estimate 
from a rough fit to a weak H$\alpha$ feature \citep*{ber97}, are 
not consistent with the astrometric distance.  As can be seen in the
Table, the relatively low spectroscopic log $g$ value makes G77-50
appear {\em underluminous}; this is atypical as overluminosity is
the signpost of binarity.  This discrepancy is resolved by presuming
the H$\alpha$ profile examined by \citet*{ber97} is due to two
velocity-shifted lines (indeed the line profile shown in Figure
23 of \citet*{ber97} appears asymmetric) or perhaps weak magnetism
which also causes Balmer line broadening and mutation in cool white
dwarfs where the lines are already weak \citep*{zuc03,ber01,ber97}. 
Its binarity has been almost certainly confirmed via the detection
of two Ca lines with disparate velocities, observed at two epochs
with both lines revealing individual velocity varation \citep*{zuc03}. 
Therefore, G77-50 is a double degenerate in which both components
are metal-rich, one or both of which have detectable hydrogen lines.
For equally luminous components, models predict $M=0.93$ $M_{\odot}$
(log $g=8.53$), and $M_V=15.74$ mag for $T_{\rm eff}=5200$ K hydrogen
atmosphere white dwarfs.  Since both components are polluted with
metals, only accretion from circumbinary or interstellar material
is consistent with the observations and a circumstellar origin can
be ruled out.  \citet*{koe05} lists this system as helium-rich
based on the absence of H$\beta$ (D. Koester 2007, private
communication)

{\em EC 1124$-$293}.  The parallax reported in \citet*{ber01} is 
an unpublished trigonometric measurement (M. T. Ruiz 2006, private 
communication) of $\pi_{\rm trig}=0''.0164\pm0''.0017$ which implies
a very low mass white dwarf or a binary (or both), if accurate.  The 
difference in the absolute magnitudes of the spectroscopic ($M_V=12.39$ 
mag) and photometric ($M_V=11.09$ mag) parameter determinations implies
an extra source with $M_V=11.48$ mag which is obviously much brighter
than an equally luminous companion to the spectroscopically identified 
star.  Therefore, either the system contains a very low mass DZ white
dwarf with a proportionally large radius so that it dominates the binary
spectral energy distribution, or the trigonometric parallax is inaccurate.
Given the spectroscopic analysis of \citet*{koe01}, which yields a very 
normal 0.6 $M_{\odot}$ DA white dwarf, it is difficult to currently
reconcile this system as binary.  Radial velocity measurements by
\citet*{zuc03} in 1998 December and 1999 April agree within the errors
and give a gravitational redshift corrected velocity of $v_r=1$ km ${\rm
s}^{-1}$ for the H$\beta$ line (no errors are given, but are likely no
greater than a few km ${\rm s}^{-1}$), while similar measurements 
reported by \citet*{pau06,pau03} yield $v_r=-3\pm4$ km ${\rm s}^{-1}$
(no epoch given).

\citet*{zuc03} find narrow to damning evidence of binarity at some 
known or suspected double degenerates \citep*{ber01,ber97} such as: 
broad H$\beta$ cores (e.g. G141-2, Case 2), presumably from two
velocity-shifted yet unresolved cores; two separate H$\beta$ cores
(e.g. G271-115, G77-50); and a single, variable H$\beta$ core (e.g.
LHS 1549).  Two of these have been confirmed as binary by other 
methods: G141-2 has been spatially resolved with the {\em Hubble} 
Fine Guidance Sensors (E. P. Nelan 2006, private communication) and
radial velocity monitoring of LHS 1549 has determined its orbital
period \citep*{nel05}.

\subsection{PG 0235$+$064}

The IRAC photometry of this target was problematic due to a nearby
M dwarf common proper motion companion which is reported here for 
the first time.  The companion, PG 0235$+$064B, is separated from the
white dwarf primary by $7''.4$ at $344\arcdeg$ in the IRAC 3.6 $\mu$m 
image (epoch 2005.6), shown in Figure \ref{fig10}.  Examining archival
images from 2MASS, and the DSS reveals separations and position angles 
between the A and B components which remain essentially constant between 
1950 and 2001.  Blinking the 1950 and 1990 DSS frames clearly shows the 
pair moving together over 40 yr.  The USNO-B1.0 catalog has $\mu=0''.18$ 
${\rm yr}^{-1}$ at $\theta=153\arcdeg$ for the white dwarf \citep*{mon03}, 
which would have caused the pair to separate by almost $10''$ over the 
last 55 yr if the secondary were background.  Hence the pair is bound.  

\begin{figure}
\plotone{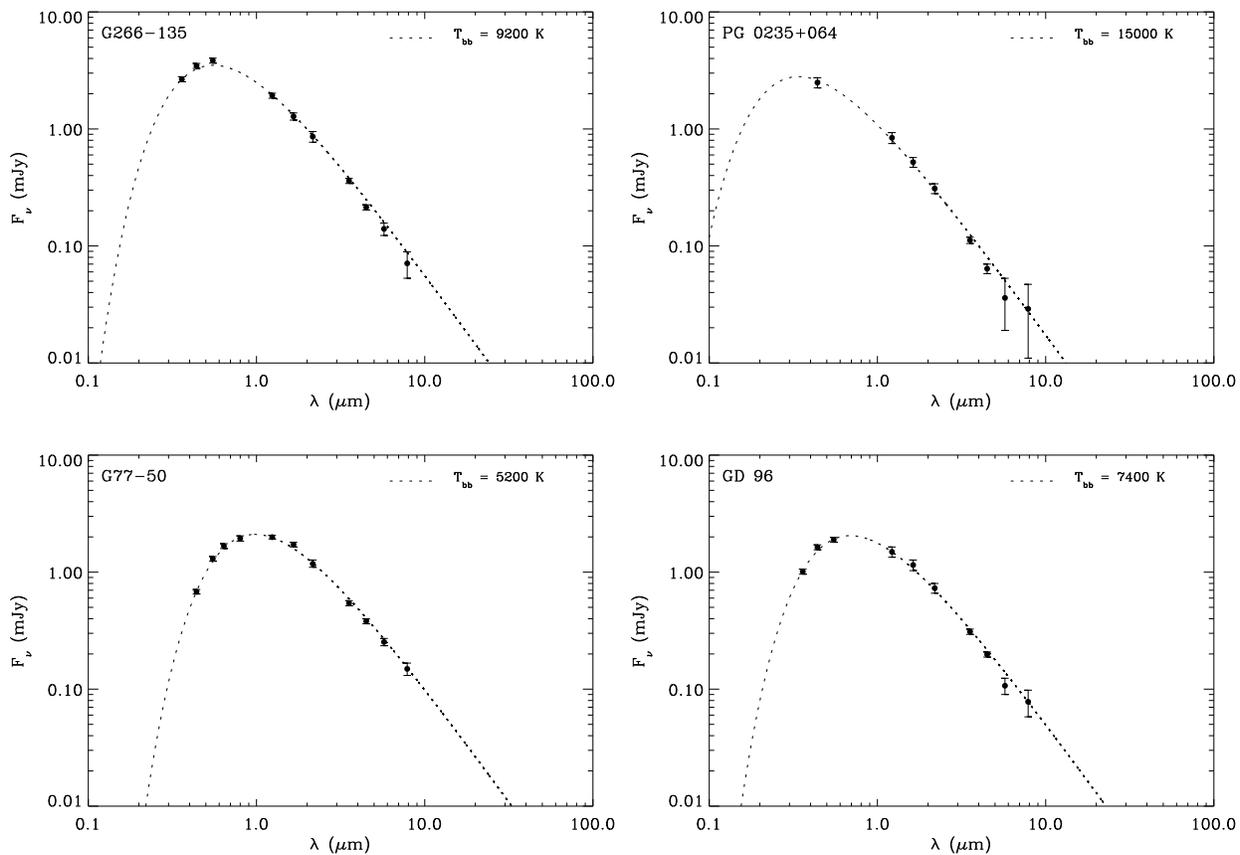}
\caption{Spectral energy distribution of G266-135, PG 0235$+$064,
G77-50, and GD 96 (see \S4.1).  For PG 0235$+$064, the light of its 
nearby M dwarf companion has been removed (see \S4.6 and Figure 
\ref{fig10}), and a superior fit to the data is achieved for an effective 
temperature of 15,000 K, contrary to previous, significantly lower 
estimates.
\label{fig1}}
\end{figure}

\begin{figure}
\plotone{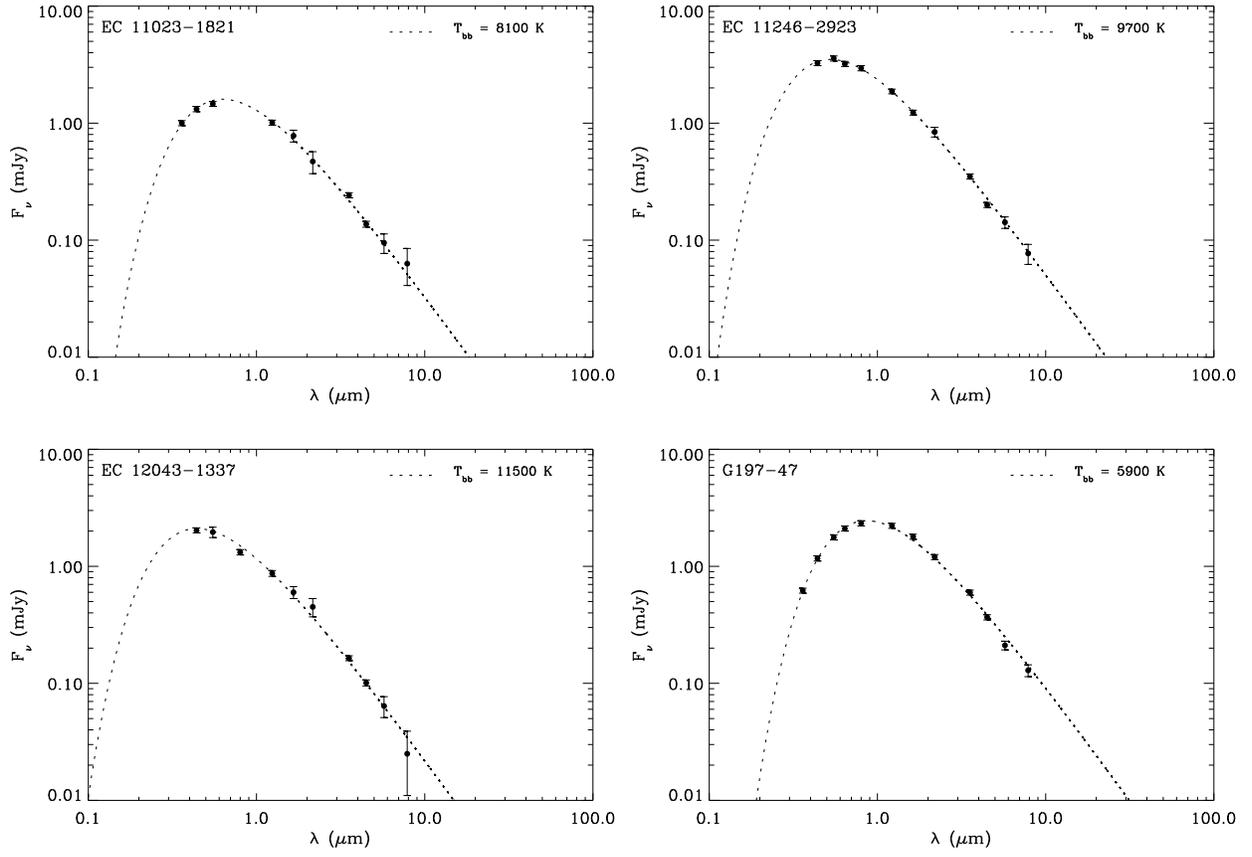}
\caption{Spectral energy distribution of EC 1102$-$183, EC 1124$-$293,
EC 1204$-$136, and G197-47.
\label{fig2}}
\end{figure}

\begin{figure}
\plotone{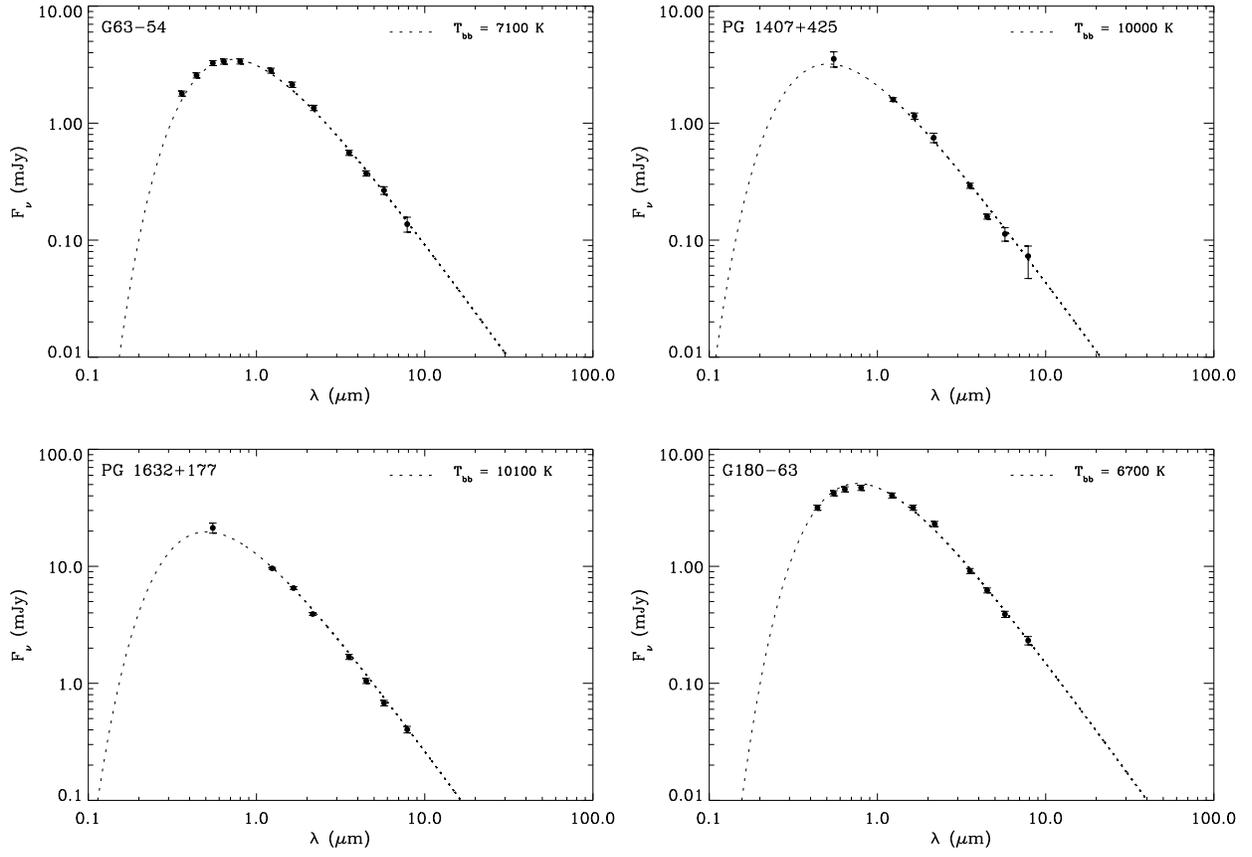}
\caption{Spectral energy distribution of G63-54, PG 1407$+$425,
PG 1632$+$177, and G180-63.
\label{fig3}}
\end{figure}

\begin{figure}
\plotone{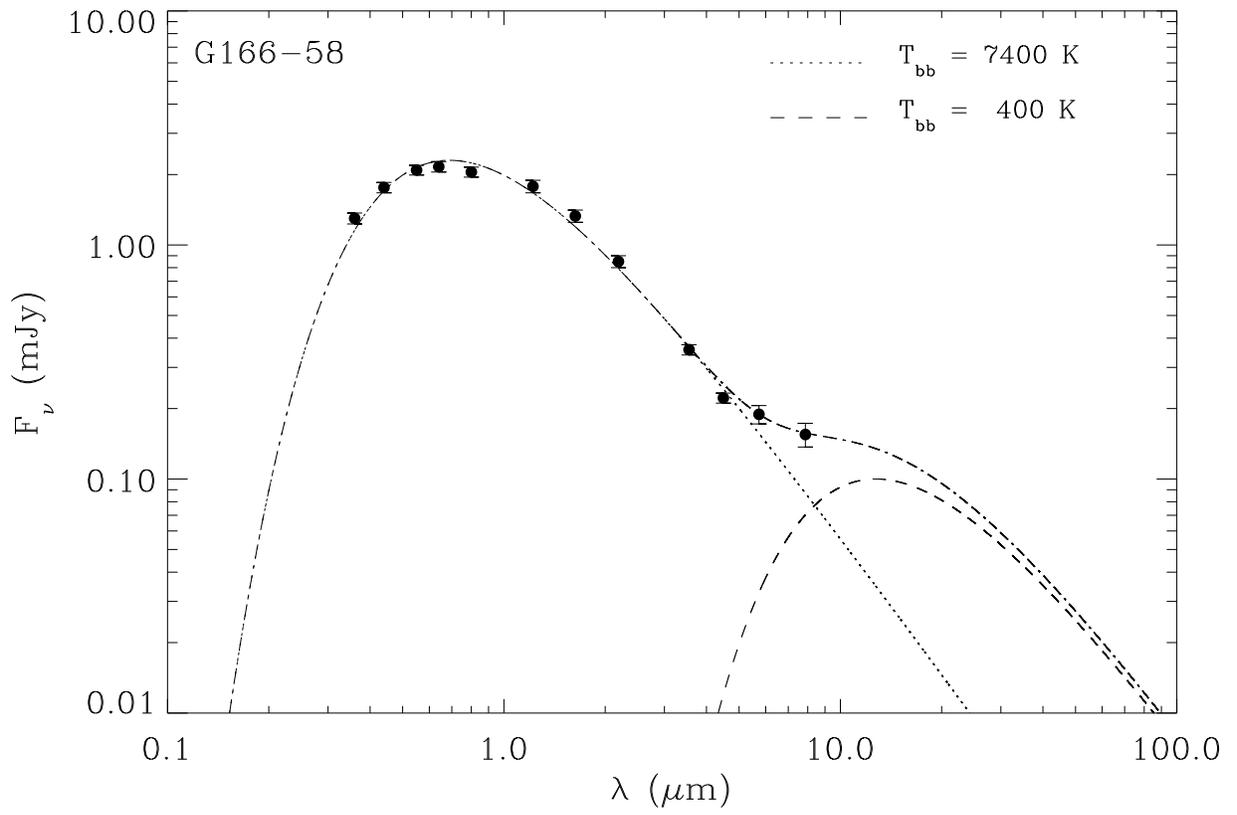}
\caption{Spectral energy distribution of G166-58.
\label{fig4}}
\end{figure}

\begin{figure}
\plotone{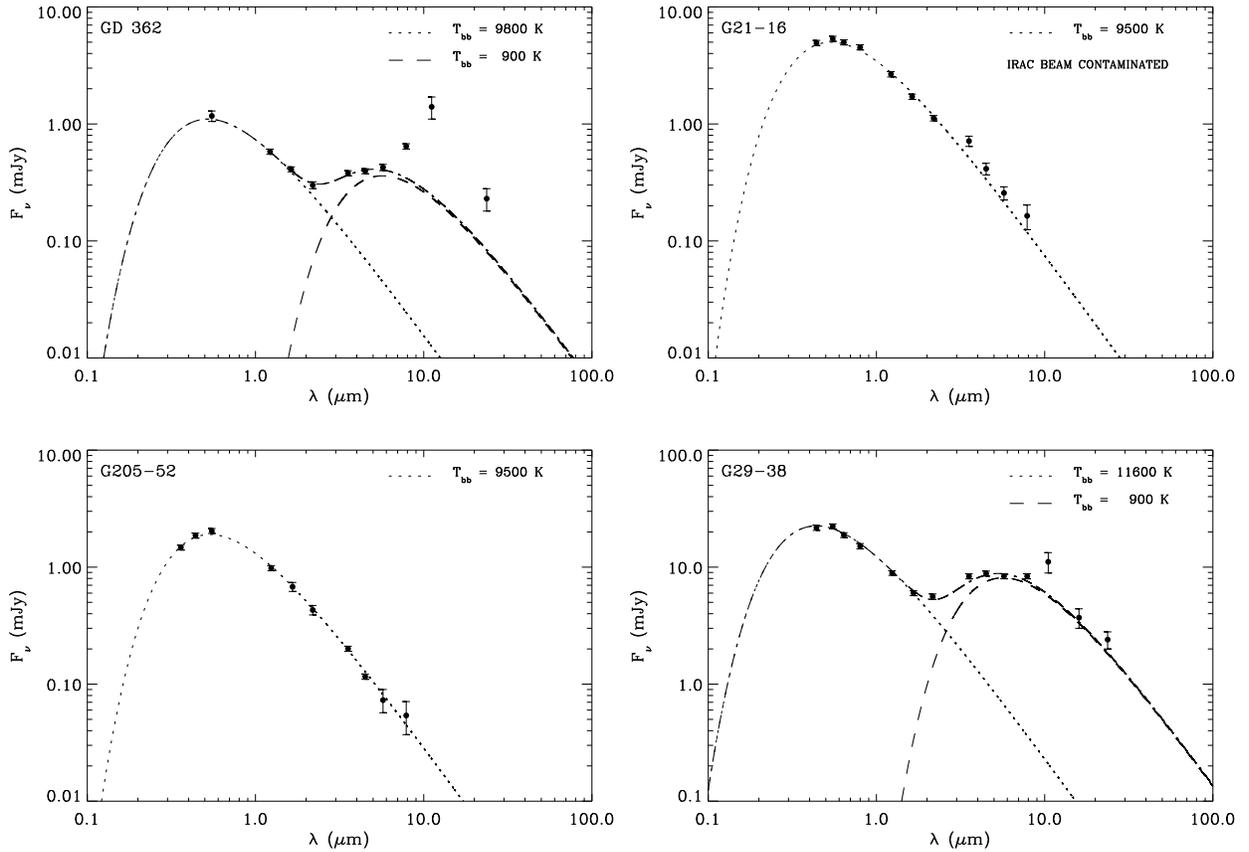}
\caption{Spectral energy distribution of GD 362, G21-16,
G205-52, and G29-38.  The IRAC photometry of G21-16 is 
contaminated at all wavelengths.
\label{fig5}}
\end{figure}

\begin{figure}
\plotone{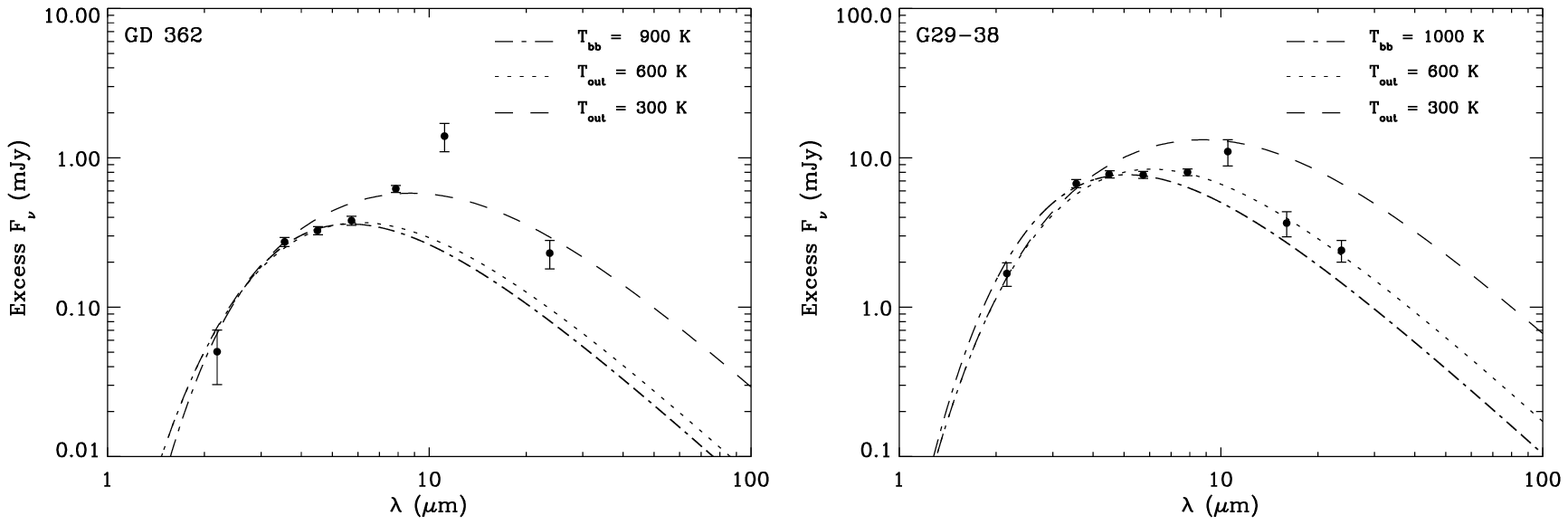}
\caption{Infrared excess flux from GD 362 and G29-38 after 
subtraction of the expected white dwarf photospheric contributions
at these wavelengths.  The data are $K$-band \citep*{skr06,bec05}, 
IRAC $3-8$ $\mu$m (this work), IRS 16 $\mu$m \citep*{rea05a}, and 
MIPS 24 $\mu$m \citep*{jur07a,rea05a}.  Plotted are models from
\citet*{jur03} for disks with an inner temperature of 1200 K and 
outer temperatures of 600 K (dotted lines) and 300 K (dashed lines) 
respectively.  Also shown are 900 K and 1000 K blackbody curves 
(dashed-dotted lines).
\label{fig6}}
\end{figure}

\begin{figure}
\plotone{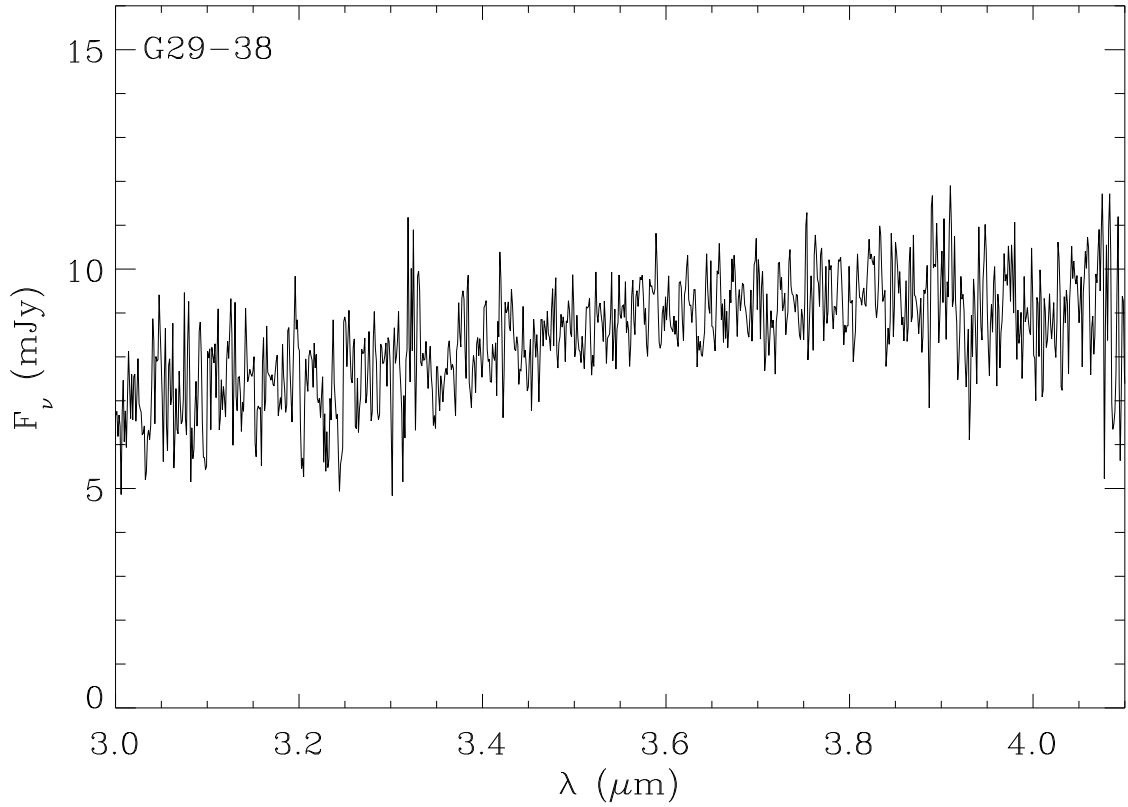}
\caption{$L$-grism spectrum of G29-38 taken with NIRI, normalized,
then flux calibrated using the IRAC 3.6 $\mu$m photometry.  There 
appears to be a gentle slope towards 4 $\mu$m, but the spectrum is
otherwise featureless.  The resolution is $\lambda/\Delta\lambda
\approx500$ and the data are neither binned nor smoothed.
\label{fig7}}
\end{figure}

\begin{figure}
\plotone{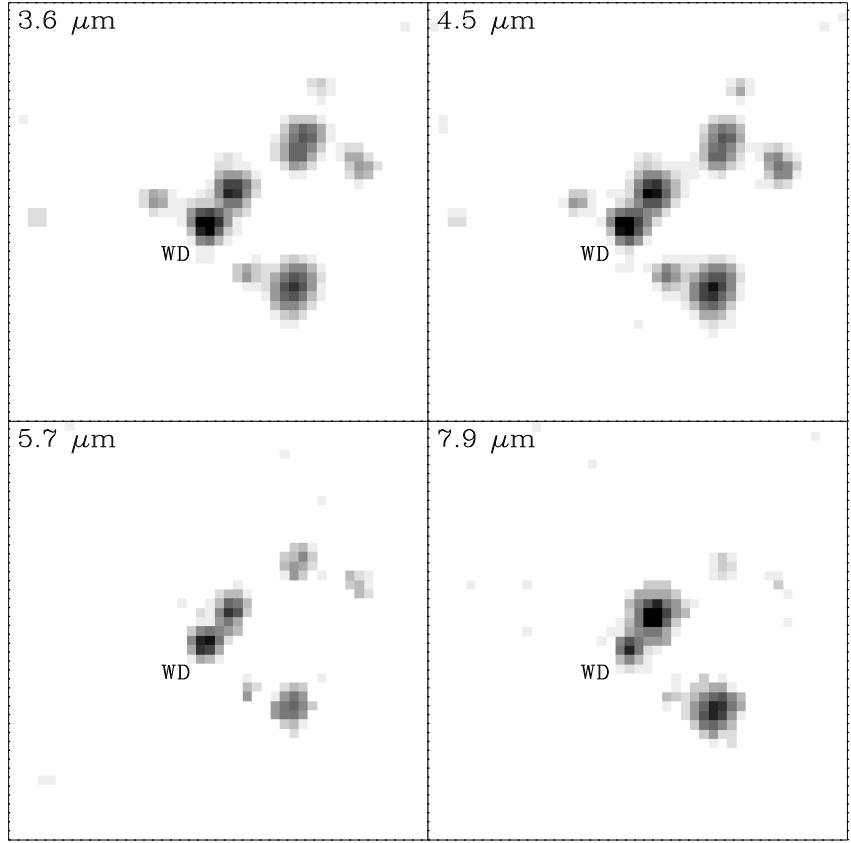}
\caption{IRAC images of G166-58 in all four channels, each spanning 
a $54''\times54''$ field of view.  The top of each image corresponds 
to position angle $125.7\arcdeg$, increasing counterclockwise.  The 
adjacent extragalactic source (discussed in \S4.4.1) is located $5''.3$
from the white dwarf at position angle $87.6\arcdeg$.  The proper motion
of G166-58 ($\mu=0.''64$ ${\rm yr}^{-1}$ at $\theta=165.2\arcdeg$; 
\citealt*{bak02}), is nearly perpendicular to the direction toward
the background source.
\label{fig8}}
\end{figure}

\begin{figure}
\plotone{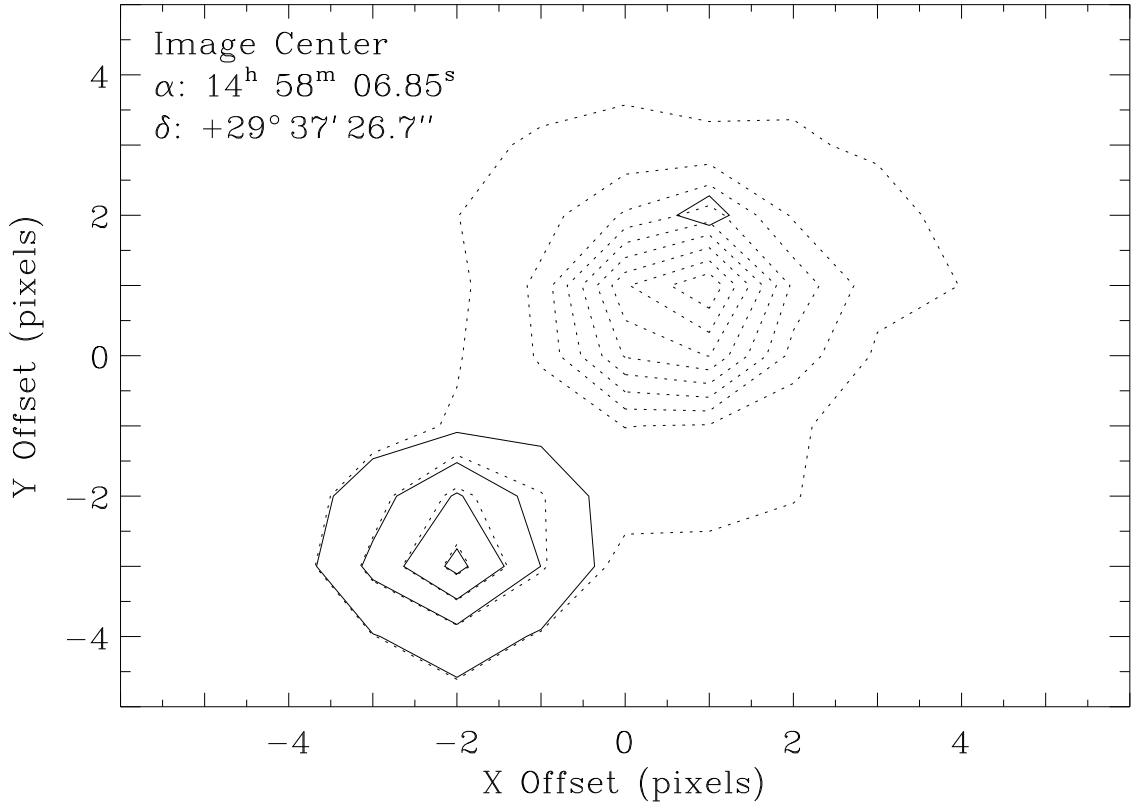}
\caption{Linear contours for 7.9 $\mu$m images of G166-58, drawn
from 1.45 to 2.80 MJy ${\rm sr^{-1}}$ in steps of 0.15 MJyr ${\rm sr^{-1}}$.  
The dotted lines show the contours for the central part of the image shown in 
the lower right panel of Figure \ref{fig8}, which includes the nearby galaxy.  
The solid lines show identical contours for a similar image where the galaxy 
has been fitted and removed as described in \S4.4.1.  The solid contour
near offset (1,2) is residual signal after subtraction of the galaxy.
\label{fig9}}
\end{figure}

\begin{figure}
\plotone{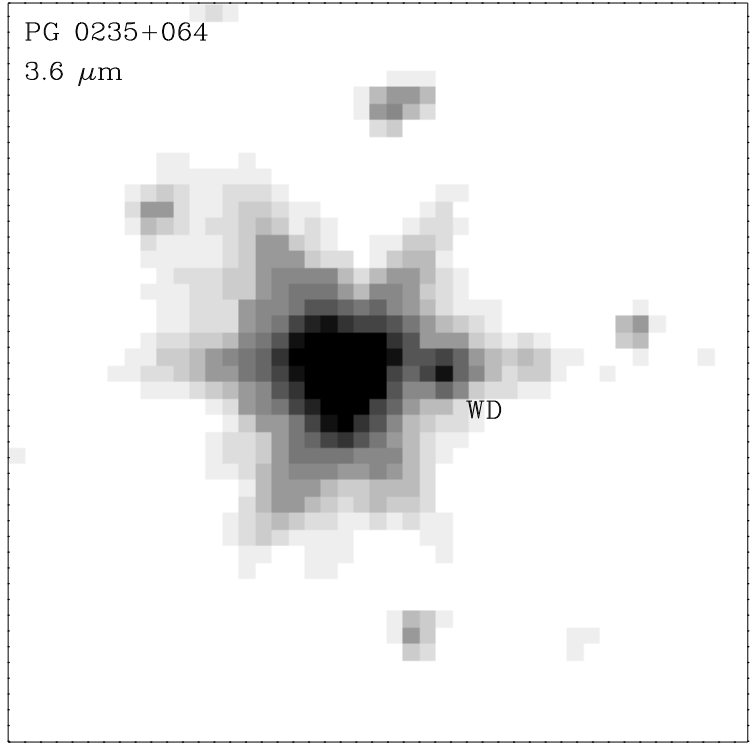}
\caption{IRAC 3.6$\mu$m image of PG 0235+064 and its M dwarf companion,
separated by $7''.4$ at $344\arcdeg$.  The top of the image corresponds
to position angle $253.3\arcdeg$, increasing counterclockwise.  The white
dwarf is quite faint at $m_{3.6\mu{\rm m}}=16.0$ mag, while the red dwarf
dominates with $m_{3.6\mu {\rm m}}=10.7$ mag.
\label{fig10}}
\end{figure}

PG 0235$+$064B has reliable 2MASS photometry consistent with an early
M dwarf which has very likely contaminated some previously published
photometry and spectroscopy of the white dwarf, even as far as the blue
optical region, causing it to appear too cool (red) \citep*{zuc03,gre86}.
Figure \ref{fig1} shows $T_{\rm eff}=15,000$ K (DA3.4) provides a better 
fit to the white dwarf data \citep*{ber95b} than previously published, lower
effective temperatures corresponding to DA4.4 and DA8 \citep*{hom98,gre86}.
The shorter wavelength photometric data in Figure \ref{fig1} were selected 
as to be minimally, or not at all contaminated by the cool companion: $B$ 
from \citet*{gre86}; $J$ from \citet*{skr06}; and $HK$ from \citet*{kil06}.
Assuming log $g=8$, very close to that determined by \citet*{hom98} from a
spectrum likely to be contaminated by the M dwarf, the white dwarf would 
lie at $d=70$ pc.  Using the {\em Hubble} GSC2.2 \citep*{sts01} blue and
red magnitudes, one can estimate $B\approx17$ mag and $B-K\approx6$ for 
the red dwarf.  This corresponds to a spectral type near M3 and agrees 
reasonably well with the expected absolute magnitude of $M_K=6.7$ mag 
at the estimated white dwarf distance \citep*{kir94}.  

For all of the IRAC images, the task {\sf daophot} was used in an
attempt to remove the light of the M dwarf, but it was found that this 
generally oversubtracted its flux in the region of the white dwarf, and 
combined with the $5-6$ mag difference in brightness between components, 
proved unreliable.  Instead, the symmetry of the point spread function
was exploited to self-subtract the flux on the opposite side of the M 
dwarf at the location of the white dwarf, introducing an additional 
error component equal to the square root of the percent flux removed.

\section{DISCUSSION AND CONCLUSIONS}

Although a survey of only 17 stars does not allow robust statistics,
it is clear that the majority of DAZ white dwarfs do not harbor warm
dust of sufficient emitting surface area to be detected with IRAC.
If most or all of these stars do host circumstellar material, the
fractional luminosities must be relatively low compared to currently
known dusty white dwarfs.  This could result from a modest amount of
dust (as in the zodiacal cloud), large particle sizes, or cooler 
material further from the star.

Another possibility is that any warm dust produced within the Roche
limit of a white dwarf is swiftly destroyed through mutual collisions,
not unlike ice in a blender, as it orbits with Keplerian velocities near 
$0.003c$.  In optically thin disks, particles with orbital period $p$ will
collide on a timescale given by $t_{\rm coll}=p/\tau$.  The tidal rings at
G29-38 and GD 362 have been modeled to extend from approximately $0.2-0.4$
$R_{\odot}$, where a typical orbital period is only $p=0.6$ hr and the
resulting collision timescale for $\tau\sim0.01$ is $t_{\rm coll}=2.5$ dy.
If a sizeable fraction of dust produced in a tidal disruption event is 
initially optically thin, then both collisions and Poynting-Robertson 
drag will compete to quickly annihilate this material.  The ratio of 
these two timescales for dust particles orbiting a distance $D$ from
a star of mass $M$ can be written as

\begin{equation}
\gamma  = \frac{t_{\rm pr}}{t_{\rm coll}} 
	= 693 \ \frac{\sqrt{MD} \rho a\tau}{LQ_{\rm pr}}
\end{equation}

\noindent
where $M$ and $L$ are in solar units, $D$ is in AU, $\rho$ is in g
${\rm cm}^{-3}$ , and $a$ is in microns.  This fraction reaches a minimum
for 0.1 $\mu$m grains with 1 g ${\rm cm}^{-3}$ at the inner disk edge, and
yields $\gamma_{\rm min}>10$ for all possible white dwarf disk parameters.
Table \ref{tbl6} gives minimum values for $\gamma$ at the inner edges of 
the disks at G29-38, GD 362, and G166-58.  Therefore collisions will erode 
dust grains faster than they can be removed by angular momentum loss.  
This is also true in the case of optically thick disks where: 1) the bulk
of material is shielded from starlight, and hence the Poynting-Robertson
effect is diminished, and 2) the collision timescale is less than half 
the orbital period \citep*{esp93}.  Hence, for a wide range of disk 
densities, it is plausible that mutual collisions within an evolving 
dust ring at a typical white dwarf will result in the relatively rapid
self-annihilation of the micron size grains required to radiate 
efficiently at $3-30$ $\mu$m.

Following the tidal disruption of an asteroid, if one models the dust
produced as a collisional cascade, the expected particle size distribution
behaves classically as $n(a)\propto a^{-3.5}$ \citep*{doh69}.  For dust at
main sequence stars, this distribution is reshaped on short timescales as 
sub-micron size grains are removed by radiation pressure which, as shown 
above, does not apply in the case of white dwarfs.  In the absence of 
radiation pressure, the average particle in will have a size $\bar{a}=
5/3$ $a_{\rm min}$.  For practical purposes, at white dwarfs one can 
assume $a_{\rm min}=0.01$ $\mu$m, where particles are already inefficient
absorbers and emitters of infrared radiation, and anything smaller approaches
the size of gas molecules and atoms.  With such a distribution of extant dust,
99.7\% of the particles will have sizes $a\leq0.1$ $\mu$m, leaving a paltry 
fraction of larger particles which could effectively support infrared emission 
from the disk.  Clearly, such small particles and gas might be present at most
or all white dwarfs which show signs of circumstellar accretion such as the
DAZ stars.

On the other hand, the persistence of warm dust disks at several white
dwarfs \citep*{jur07b} must be explained despite the fact that in some
or most cases where it is produced, it may also be efficiently destroyed.
One possibility is that the disk density (which could contain gas) becomes
sufficiently high as to damp out collisions in the disk and also make it
optically thick, thus somewhat protecting it from self-erosion and drag
forces simultaneously.  The evolution of such a dense, fluid-like ring is
then dominated by viscous forces (differential rotation and random motions)
which cause it to spread, losing energy in the process \citep*{esp93}.  The
maximum lifetime of such a ring occurs at minimum viscosity

\begin{equation}
t_{\rm ring} \approx \frac{ {{\rho}^2} {w^2} p} {2 \pi {{\sigma^2}} }
\end{equation}

\noindent
where $w$ is the radial extent of the ring and $\sigma$ is the surface 
mass density \citep*{esp93}.  If the mass of a large solar system asteroid, 
$10^{24}$ g, were spread into a tidal ring of negligible height ($h<10$ m),
a radial extent $0.2-0.4$ $R_{\odot}$, consisting of micron size particles
orbiting a typical white dwarf, the resulting volume mass density (0.55 g 
${\rm cm}^{-3}$) would be sufficiently high that the mean free path of 
particles is on the same order as their size.  This could effectively damp
out collisions, thus minimizing viscosity, and with a resulting surface
mass density of $\sigma\approx550$ g ${\rm cm}^{-2}$, permit a potential 
disk lifetime -- in the absence of competing forces -- longer than the
Gyr white dwarf cooling timescales.  However, for sustained accretion
rates as low as $\dot{M}=10^{10}$ g ${\rm s}^{-1}$, a $10^{24}$ g disk
would become fully consumed within several Myr.

Large rocks and colder material orbiting at $D\ga100$ $R_{\odot}$ will be
unaffected by any of the aforementioned processes, and such a reservoir of
material is strictly necessary to supply some fraction of DAZ white dwarfs
with photospheric metals, regardless of circumstellar dust production 
(collisional versus tidal) and evolution (persistence versus destruction).

The overall number of white dwarfs with remnant planetesimal belts
may be rather high based on a growing number of detections.  If one 
takes 12\% (\S4.1) as the fraction of DAZ stars with circumstellar dust as
observed by {\em Spitzer} to date, 20\% as the fraction of DAZ stars among 
cool DA white dwarfs \citep*{zuc03}, and 80\% as the number of cool DA stars
among all white dwarfs in the field \citep*{eis06}, then a lower limit to 
the number of white dwarfs with asteroid-type belts is at least 2\%.  This
fraction could be as high as 20\% if the majority of metal-rich white dwarfs
harbor circumstellar matter, which raises important questions about the implied
frequency of planetesimal belts around main-sequence stars and the current 
detection rate (see the Appendix of \citealt*{jur06}).

Owing to their low luminosities, white dwarfs which may have been 
polluted by heavy elements in winds or transferred material from 
substellar companions \citep*{deb06,dob05,zuc03,sio84} are easily 
identified with IRAC observations \citep*{mul07,han06,far05a,far05b} 
down to T dwarf temperatures.  There is no evidence of such companions 
in the data presented here, ruling out all but the coldest brown dwarfs,
active planets and moons as close orbiting, companion-like polluters 
(Farihi et al. 2008, in preparation).

\acknowledgments

J. Farihi thanks M. Jura for helpful discussions on circumstellar dust,
S. Fisher for his expertise on mid-infrared detectors and photometry, T. 
Geballe for encyclopedic assistance with $3-4$ $\mu$m spectra, S. Wachter
and S. Carey for sharing their familiarity with {\em Spitzer} instruments
and data, and P. Bergeron for kindly providing access to current white 
dwarf models.  This work is based on observations made with the {\em 
Spitzer Space Telescope}, which is operated by the Jet Propulsion 
Laboratory, California Institute of Technology under a contract with 
NASA.  Support for this work was provided by NASA through an award 
issued by JPL/Caltech.  Spectroscopic observations for this work were
taken as part of the Gemini Director's Discretionary Time GN-2005B-DD-1.
Gemini Observatory is operated by the Association of Universities for
Research in Astronomy, Inc., under a cooperative agreement with the NSF
on behalf of the Gemini partnership: the National Science Foundation
(United States), the Particle Physics and Astronomy Research Council 
(United Kingdom), the National Research Council (Canada), CONICYT 
(Chile), the Australian Research Council (Australia), CNPq (Brazil),
and CONICET (Argentina).

{\em Facility:} \facility{Spitzer (IRAC)}, \facility{Gemini (NIRI)}

\clearpage

\begin{deluxetable}{cccccc}
\tablecaption{DAZ\tablenotemark{a} White Dwarf Targets\label{tbl1}}
\tablewidth{0pt}
\tablehead{
\colhead{WD}			&
\colhead{Name}			&
\colhead{$T_{\rm eff}$ (K)}	&
\colhead{$V$ (mag)}			&
\colhead{[Ca/H]}			&
\colhead{References}}

\startdata

0032$-$175	&G266-135	&9240	&14.94	&-10.20	&1,2\\
0235$+$064	&PG			&15000	&15.5	&-9.03	&1,3\\
0322$-$019	&G77-50		&5220	&16.12	&-11.36	&1,4\\
0846$+$346	&GD 96		&7370	&15.71	&-9.41	&1,5\\
1102$-$183	&EC			&8060	&15.99	&-10.43	&1,5\\
1124$-$293	&EC			&9680	&15.02	&-8.53	&1,4,6\\
1204$-$136	&EC			&11500	&15.67	&-7.72	&1,7\\
1208$+$576	&G197-47		&5880	&15.78	&-10.96	&1,8\\
1344$+$106	&G63-54		&7110	&15.12	&-11.13	&1,8\\
1407$+$425	&PG			&10010	&15.03	&-9.87	&1,9\\
1455$+$298	&G166-58		&7390	&15.60	&-9.31	&1,8,9\\
1632$+$177	&PG			&10100	&13.05	&-10.75	&1,9\\
1633$+$433	&G180-63		&6690	&14.84	&-8.63	&1,8,9\\
1729$+$371	&GD 362		&10500	&16.23	&-5.1	&7,10\\
1826$-$045	&G21-16		&9480	&14.58	&-8.83	&1,8\\
1858$+$393	&G205-52		&9470	&15.63	&-7.84	&1,5\\
2326$+$049	&G29-38		&11600	&13.04	&-6.93	&1,5\\

\enddata

\tablenotetext{a}{G77-50 and GD 362 are helium-rich (\citealt*{koe05};
D. Koester 2007, private communication}

\tablerefs{
(1) \citealt*{zuc03};
(2) \citealt*{mer86};
(3) This work;
(4) \citealt*{ber97};
(5) \citealt*{mcc03};
(6) \citealt*{koe01};
(7) \citealt*{sal03};
(8) \citealt*{ber01};
(9) \citealt*{lie05};
(10) D. Koester 2007, private communication)}

\end{deluxetable}

\clearpage

\begin{deluxetable}{cccccc}
\tablecaption{IRAC Fluxes for White Dwarf Targets\label{tbl2}}
\tablewidth{0pt}
\tablehead{
\colhead{WD}					&
\colhead{$F_{3.6\mu{\rm m}}$ ($\mu$Jy)}	&
\colhead{$F_{4.5\mu{\rm m}}$ ($\mu$Jy)}	&
\colhead{$F_{5.7\mu{\rm m}}$ ($\mu$Jy)}	&
\colhead{$F_{7.9\mu{\rm m}}$ ($\mu$Jy)}	&
\colhead{Pipeline}}

\startdata

0032$-$175	&$360\pm18$		&$214\pm11$		&$140\pm17$		&$71\pm18$		&11.0\\
0235$+$064	&$112\pm7$		&$64\pm6$		&$36\pm17$		&$29\pm18$		&12.4\\
0322$-$019	&$543\pm27$		&$381\pm19$		&$253\pm18$		&$149\pm18$		&12.4\\
0846$+$346	&$310\pm16$		&$198\pm10$		&$107\pm17$		&$78\pm20$		&10.5\\
1102$-$183	&$242\pm12$		&$137\pm8$		&$95\pm18$		&$63\pm22$		&11.0\\
1124$-$293	&$350\pm18$		&$200\pm10$		&$142\pm16$		&$77\pm15$		&12.4\\
1204$-$136	&$164\pm8$		&$101\pm6$		&$64\pm13$		&$25\pm14$		&12.4\\
1208$+$576	&$597\pm30$		&$367\pm19$		&$211\pm18$		&$129\pm15$		&11.0\\
1344$+$106	&$558\pm28$		&$372\pm19$		&$266\pm20$		&$137\pm20$		&12.4\\
1407$+$425	&$292\pm15$		&$159\pm8$		&$113\pm15$		&$73\pm16$		&12.4\\
1455$+$298	&$357\pm18$		&$222\pm11$		&$189\pm17$		&$155\pm18$		&12.4\\
1632$+$177	&$1683\pm84$		&$1049\pm53$		&$679\pm37$		&$403\pm26$		&11.4\\
1633$+$433	&$912\pm46$		&$623\pm31$		&$389\pm24$		&$232\pm19$		&11.4\\
1729$+$371	&$380\pm19$		&$395\pm20$		&$425\pm26$ 		&$644\pm34$		&12.4\\
1826$-$045	&$714\pm72$		&$414\pm47$		&$257\pm33$		&$164\pm39$		&12.4\\
1858$+$393	&$201\pm10$		&$116\pm6$		&$73\pm17$		&$54\pm17$		&10.5\\
2326$+$049	&$8350\pm420$	&$8810\pm440$	&$8370\pm420$	&$8370\pm420$	&14.0\\

\enddata

\tablecomments{Error calculations, including both photometric
measurements and instrumental uncertainties are described in \S3.2.}

\end{deluxetable}

\clearpage

\begin{deluxetable}{ccccc}
\tablecaption{Flux Determinations for G166-58\label{tbl3}}
\tablewidth{0pt}
\tablehead{
\colhead{Method}						&
\colhead{$F_{3.6\mu{\rm m}}$ ($\mu$Jy)}		&
\colhead{$F_{4.5\mu{\rm m}}$ ($\mu$Jy)}		&
\colhead{$F_{5.7\mu{\rm m}}$ ($\mu$Jy)}		&
\colhead{$F_{7.9\mu{\rm m}}$ ($\mu$Jy)}}

\startdata

1\tablenotemark{a}	&$357\pm18$		&$222\pm11$		&$189\pm17$		&$155\pm18$\\
2\tablenotemark{b}	&$410\pm23$		&$236\pm11$		&$172\pm13$		&$164\pm20$\\
3\tablenotemark{c}	&$350\pm35$		&$225\pm23$		&$170\pm17$		&$150\pm15$\\

\enddata

\tablecomments{See \S4.4.1 for a detailed description of the methods used
to account for and eliminate any and all flux from the nearby galaxy.}

\tablenotetext{a}{Aperture photometry.}
\tablenotetext{b}{Point spread function fitting ({\sf daophot}) photometry.}
\tablenotetext{c}{Radial profile analysis.}

\end{deluxetable}

\clearpage

\begin{deluxetable}{ccccc}
\tablecaption{Possible Disk Parameters for G166-58\label{tbl4}}
\tablewidth{0pt}
\tablehead{
\colhead{Disk Type}						&
\colhead{log $g$}						&
\colhead{$R_{\rm eff}$ ($R_{\odot}$)}		&
\colhead{$D_{\rm in, thick}$ ($R_{\odot}$)}	&
\colhead{$D_{\rm in, thin}$ ($R_{\odot}$)}}

\startdata

Circumstellar	&7.97	&0.0132	&0.38				&2.3\\
Circumbinary	&7.62	&0.0163	&0.47\tablenotemark{a}	&2.7\\

\enddata

\tablecomments{In the case of a wide binary where dust orbits a 
single, normal mass, white dwarf component, the dust is circumstellar.
A single, low mass white dwarf of large radius is ruled out by 
spectroscopy \citep*{lie05}.}

\tablenotetext{a}{A close binary surrounding by a flat optically
thick disk is not physically possible (\S4.4.4).}

\end{deluxetable}

\clearpage

\begin{deluxetable}{ccccccccccc}
\rotate
\tablecaption{Confirmed or Suspected Double Degenerate DAZ Systems\label{tbl5}}
\tablewidth{0pt}
\tablehead{

&&	&Spectroscopic\tablenotemark{a}
&&&	&Photometric\tablenotemark{b}&	&&\\

\cline{3-5}
\cline{7-9}
\\

\colhead{Star}				&

\colhead{}					&

\colhead{$T_{\rm eff}$ (K)}	&
\colhead{log $g$}			&
\colhead{$d$ (pc)}			&

\colhead{}					&

\colhead{$T_{\rm eff}$ (K)}	&
\colhead{log $g$}			&
\colhead{$d$ (pc)}			&

\colhead{}					&

\colhead{References}}

\startdata

G77-50			&&5220	&7.5		&23.3	&&5200	&8.04	&16.8	&&1,2,3\\
EC 1124$-$293	&&9550	&8.04	&33.6	&&9440	&7.10	&61.1	&&4,5\\
G166-58			&&7390	&7.97	&29.1	&&7340	&7.62	&35.5	&&4,6,7\\

\enddata

\tablenotetext{a}{Parameters derived from model fits to spectroscopic
continuum flux and Balmer line profiles.}

\tablenotetext{b}{Parameters derived from model fits to photometric
fluxes and trigonometric parallax.}

\tablerefs{
(1) \citealt*{ber97};
(2) \citealt*{leg98};
(3) \citealt*{sma03};
(4) \citealt*{ber01};
(5) \citealt*{koe01};
(6) \citealt*{zuc03};
(7) \citealt*{lie05}
}

\end{deluxetable}

\clearpage

\begin{deluxetable}{cccccc}
\tablecaption{Minimum $\gamma$ Values at Inner Disk Edges\label{tbl6}}
\tablewidth{0pt}
\tablehead{
\colhead{Star}						&
\colhead{$\tau$\tablenotemark{a}}		&
\colhead{$M$ ($M_{\odot}$)}			&
\colhead{log($L/L_{\odot}$)}			&
\colhead{$D_{\rm in}$ ($R_{\odot}$)}	&
\colhead{$\gamma_{\rm min}$}}

\startdata

G29-38					&0.030	&0.69	&$-2.64$	&0.14	&19\\
G166-58\tablenotemark{b}	&0.0023	&0.58	&$-3.36$	&0.38	&12\\
G166-58\tablenotemark{c}	&0.0023	&1.2		&$-3.18$	&2.7		&30\\
GD 362					&0.030	&0.75	&$-2.90$	&0.12	&34\\

\enddata

\tablecomments{The minimum value of $\gamma$ is achieved, realistically,
at $\rho=1$ g ${\rm cm}^3$ and $a=0.1$ $\mu$m.}

\tablenotetext{a}{$L_{\rm IR}/L$.}
\tablenotetext{b}{Single white dwarf with an optically thick circumstellar disk.}
\tablenotetext{c}{Double white dwarf with an optically thin circumbinary disk.}

\end{deluxetable}

\clearpage

\clearpage

\clearpage

\clearpage

\clearpage

\clearpage

\clearpage

\clearpage

\clearpage

\clearpage

\end{document}